\documentclass[conference]{IEEEtran}
\IEEEoverridecommandlockouts
\usepackage{cite}
\usepackage{amsmath,amssymb,amsfonts}
\usepackage{amsthm}
\usepackage{graphicx}
\usepackage{textcomp}
\usepackage{xcolor}
\usepackage{algorithm}
\usepackage{algpseudocode}
\usepackage{graphicx}
\usepackage{url}
\usepackage[subject={Todo},author={Bastian}]{pdfcomment}
\usepackage[normalem]{ulem}
\usepackage{bbm}
\usepackage{bm}
\usepackage{xspace}
\usepackage{multirow}

\newcommand{\ie}[0]{i.e.,\xspace}
\newcommand{\eg}[0]{e.g.,\xspace}

\newtheorem{theorem}{Theorem}

\newtheorem{definition}[theorem]{Definition}
\newtheorem{example}[theorem]{Example}

\def\BibTeX{{\rm B\kern-.05em{\sc i\kern-.025em b}\kern-.08em
    T\kern-.1667em\lower.7ex\hbox{E}\kern-.125emX}}
\begin{document}

\title{DBA bandits: Self-driving index tuning under ad-hoc, analytical workloads with safety guarantees}

\author{
\IEEEauthorblockN{R. Malinga Perera, Bastian Oetomo, Benjamin I. P. Rubinstein,  Renata Borovica-Gajic}
\IEEEauthorblockA{\{malinga.perera, b.oetomo\}@student.unimelb.edu.au,  \{brubinstein, renata.borovica\}@unimelb.edu.au}
\IEEEauthorblockA{\textit{School of Computing and Information Systems} \\
	\textit{University of Melbourne}}}

\maketitle

\begin{abstract}
Automating physical database design 
has remained a long-term interest in database research due to substantial performance gains afforded by optimised 
structures. Despite significant progress, a majority of today's commercial solutions are highly manual, requiring offline invocation by database administrators (DBAs) who are expected to identify and supply representative training workloads. Even the latest advancements like query stores provide only limited support for dynamic environments.
This status quo is untenable: identifying \emph{representative} static workloads is no longer realistic;
and physical design tools remain susceptible to the query optimiser's cost misestimates (stemming from unrealistic assumptions such as attribute value independence and uniformity of data distribution).

We propose a self-driving approach to online index selection that eschews the DBA and query optimiser, and instead \emph{learns} the benefits of viable structures through strategic exploration and direct performance observation. We view the problem as one of sequential decision making under uncertainty, specifically within the bandit learning setting. 
Multi-armed bandits balance exploration and exploitation to \emph{provably} guarantee average performance that converges to a fixed policy that is optimal with perfect hindsight. Our comprehensive empirical results demonstrate up to 75\% speed-up on shifting and ad-hoc workloads and 28\% speed-up on static workloads compared against a state-of-the-art commercial tuning tool.
\end{abstract}

\section{Introduction}
\label{sec:intro}
With the increasing complexity and variability of database applications and their hosting platforms (\eg multi-tenant cloud environments), automated physical design tuning, particularly automated index selection, has re-emerged as a contemporary challenge for database management systems. %
Most database vendors offer automated tools for physical design tuning within their product suites~\cite{DTA2005, DB2_IntegratedApproach, OracleAdvisor}. 
Such tools 
form an integral part of broader efforts toward fully automated database management systems which aim to:
a) decrease database administration costs and thus total costs of ownership~\cite{TCO, pavlo};
b) help non-experts use database systems; and
c) facilitate hosting of databases %
on dynamic environments such as cloud-based services%
~\cite{CloudDBAsAService, sqlvm, das2019automatically, AIMeetsAI}. %
Most physical design tools take an \emph{off-line} approach, where the representative training workload is provided by the database administrator (DBA)%
~\cite{SurajitDecade}.
Where \emph{online} solutions are provided~\cite{COLT2, QUIET, OnlineApproachAutoAdmin, ToTuneOrNot, das2019automatically, forecastingPavlo}%
, questions remain: How often should the tools be invoked? 
And more importantly, is the quality of proposed designs in any way guaranteed? 
How can tools generalise beyond queries seen %
to dynamic ad-hoc workloads, where queries are unpredictable and non-stationary?

Modern analytics workloads are dynamic in nature with ad-hoc queries 
common~\cite{ResearcherGuide} \eg data exploration %
workloads adapt to past query responses~\cite{StratosExploration, noDBSIGMOD2012}. 
Such ad-hoc workloads hinder automated tuning since:
a) inputting representative information to design tools is infeasible under time-evolving workloads; and
b) %
reacting too quickly to changes may result in undesirable performance variability, where indices are continuously dropped and created. 
Any proposed robust automated physical design solution must address such challenges
~\cite{COLT2}.

To compare alternative physical design structures, automated design tools use a cost model employed by the query optimiser, typically exposed through a ``what-if" interface~\cite{ChaudhuriWhatIf}, as the sole source of truth. 
However such cost models %
make inappropriate assumptions about data characteristics~\cite{ImplicationsOfAssumptions, HowGoodQO}: 
commercial DBMSs often assume uniform data distributions and attribute value independence. %
As a result, estimated benefits of proposed designs may diverge significantly from actual workload performance~\cite{DBTest2012PhysicalDesigners, GebalyRobustness, vldbjborovica, AIMeetsAI, das2019automatically}. %
Even with more complex data distribution statistics such as single- and multi-column histograms, the issue remains for complex workloads, as demonstrated in our experiments (Section \ref{sec:evaluation}).

\begin{figure*}[t]
    \centering
    \includegraphics[width = \textwidth]{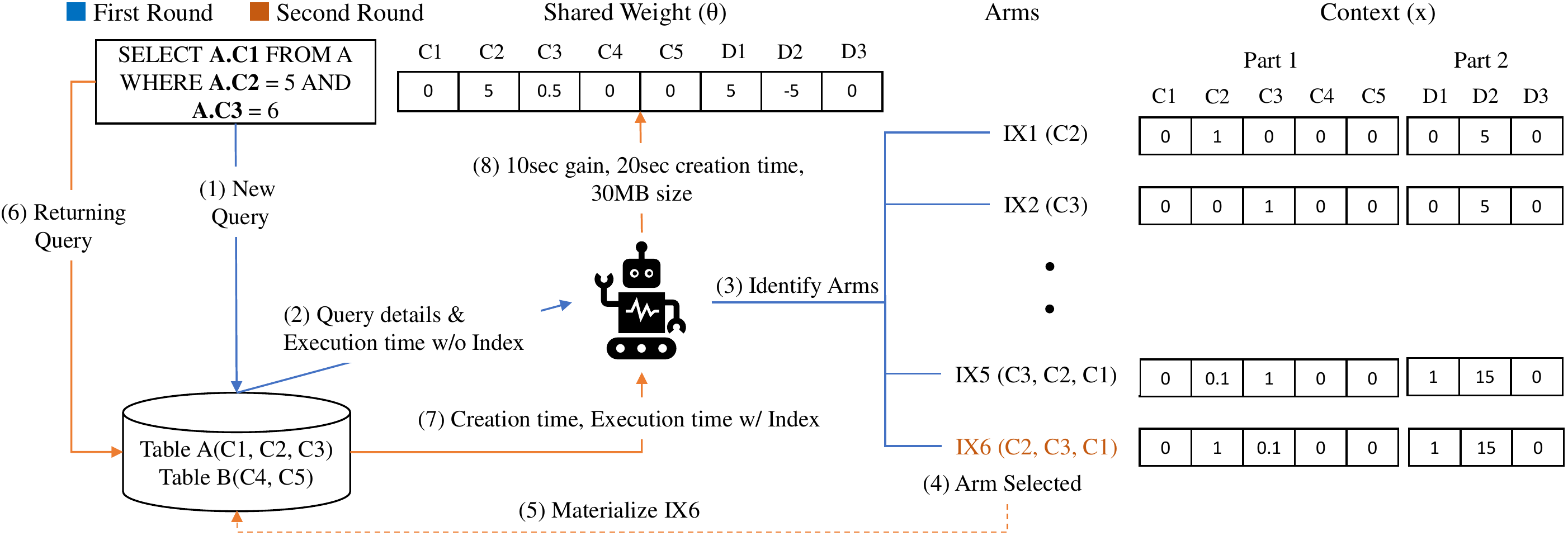}
\caption{An abstract view of the proposed bandit learning-based online index selection.}
    \label{fig:bandit_overall}
\end{figure*}

In this paper, we demonstrate that even in ad-hoc environments where queries are unpredictable, there are opportunities for index optimisation. We argue that the problem of online index selection under ad-hoc, analytical workloads can be efficiently formulated within the 
multi-armed bandit (MAB) learning setting---a tractable form of Markov decision process.
MABs take arms or actions (selecting indices) to maximise cumulative rewards, trading off exploration of untried actions with exploitation of actions that maximise rewards observed so far (see Figure~\ref{fig:bandit_overall}).  
MABs permit learning %
from observations of actual performance, and need not rely on potentially misspecified cost models. %
Unlike initial efforts with applying learning for physical design, \eg more general forms of reinforcement learning~\cite{no_dba}, bandits offer regret bounds that guarantee the fitness of dynamically-proposed indices~\cite{c2ucb}.

The key contributions of the paper are summarised next:
\begin{itemize}
    \item We model index tuning as a multi-armed bandit, proposing design choices that lead to a practical, competitive solution;
    \item Our proposed design achieves a worst-case safety guarantee
    against any optimal fixed policy, as a consequence of a corrected regret analysis of the C$^2$UCB bandit; %
    and
    \item Our comprehensive experiments demonstrate MAB's superiority over a state-of-the-art commercial physical design tool, with up to 75\% speed-ups under dynamic, analytical workloads.
\end{itemize}
\section{Problem Formulation}
\label{sec:problem_formulation}

The goal of the \emph{online database index selection problem} is to choose a set of indices (henceforth referred to as a \emph{configuration}) that minimises the total running time of a workload sequence within a given memory allowance. Neither the workload sequence, nor system run times, are known in advance.

We adopt the problem definition of~\cite{OnlineApproachAutoAdmin}. Let the \emph{workload} $W = (w_{1},w_{2},\ldots,w_{T})$ be a sequence of \emph{mini-workloads} (\eg a sequence of single queries), $I$ the set of \emph{secondary indices}, 
$C_{mem}(s)$ represent the memory space required to materialise a configuration $s\subseteq I$,
and $\mathcal{S}=\left\{s\subseteq I \left| \, C_{mem}(s) \leq M\right\} \right. \subseteq 2^I$ be the class of \emph{index configurations} feasible within our total memory allowance $M$. Our goal is to propose a configuration sequence $S = (s_{0},s_{1},\ldots,s_{T})$, with $s_t \in \mathcal{S}$ as the configuration in round $t$ and $s_{0}=\emptyset$ as the starting configuration, which minimises the \emph{total workload time $C_{tot}(W,S)$} defined as:
$$C_{tot}(W,S) = \sum_{t=1}^T C_{rec}(t) + C_{cre}(s_{t-1},s_t) + C_{exc}(w_t,s_t)\enspace.$$
Here $C_{rec}(t)$ refers to the \emph{recommendation time} in round $t$ (defined as running time of the recommendation tool) and $C_{cre}(s_{t-1},s_t)$ refers to the incremental index creation time in transitioning from configuration $s_{t-1}$ to $s_{t}$. Finally, $C_{exc}(w_t,s_t)$ denotes the execution time of mini-workload $w_{t}$ under the configuration $s_{t}$, namely the sum of response times of individual queries.

At round $t$, the system:
\begin{enumerate}
    \item Chooses a set of indices $s_{t} \in \mathcal{S}$ in preparation for upcoming workload $w_t$, without  direct access to $w_t$. 
    
    $s_t$
    only depends on observation of historical workloads ($w_1, \ldots, w_{t-1}$), corresponding sets of chosen indices, and resulting performance;
    
    \item Materialises the indices in $s_t$ which do not exist yet, that is, all indices in the set difference
    $s_t \backslash s_{t-1}$; and
    \item Receives workload $w_t$, executes all the queries therein, and measures elapsed time of each individual query and each operator in the corresponding query plan.
\end{enumerate}

\section{Contextual Combinatorial Bandits}
\label{sec:background}
In this paper, we argue that online index selection can be successfully addressed using  multi-armed bandits (MABs) from statistical machine learning, where different arms correspond to chosen indices. 
We first present necessary background on MABs, outlining the essential properties that we exploit in our work (i.e., bandit context and combinatorial arms) to converge to a highly performant index configuration. %

We use the following notation.
We denote non-scalar values with boldface: lowercase for vectors and uppercase for matrices. 
We also write $[k]=\{1,2,\ldots,k\}$ for $k\in\mathbb{N}$, and denote the transpose of a matrix or a vector with a prime.

The contextual combinatorial bandit setting under  \emph{semi-bandit} feedback involves repeated selections from $k$ possible actions, over rounds $t=1,2,\ldots$, in which the MAB:
\begin{enumerate}
    \item Observes a \emph{context} feature vector (possibly random or adversarially chosen) of each action or \emph{arm} $i \in [k]$, denoted as $\bm{X}_t = \{ \bm{x}_t(i) \}_{i \in [k]}$, for $\bm{x}_t(i) \in \mathbb{R}^d$, along with their costs, $c_i$;
    \item Selects or \emph{pulls}\ a set of arms (often referred to as \emph{super arm}) $s_t\in \mathcal{S}_t$, where we restrict the class of possible super arms $\mathcal{S}_t \subseteq \mathcal{S}^\prime_t = \left\{s \subseteq [k] \left| \, \sum_{i\in s}c_i \leq C \right. \right\} \subseteq 2^{[k]}$; and 
    \item For each $i_t \in s_t$, observes random \emph{scores} $r_t(i_t)$ drawn from fixed but unknown arm distribution which depends solely on the arm $i_t$ and its context $\bm{x}_t(i_t)$, whose true expected values are contained in the unknown variable $\bm{r}^\star_t = \{\mathbb{E}[r_{t}(i)]\}_{i \in [k]}$
\end{enumerate}
A MAB's goal is to maximise the cumulative expected reward $\sum_t \mathbb{E}[R_t(s_t)] = \sum_t g(\bm{r}^\star_t, \bm{X}_t, s_t)$ for a known function $g$. This function $g$ need not be a simple summation of all the scores. The core challenge in this problem is that the expected scores for all arms $i\in[k]$ are unknown. Refinement of a bandit learner's approximation for arm $i$ is \emph{generally} only possible by including arm $i$ in the super arm, as the score for arm $i$ is not observable when $i$ is not played. This suggests solutions that balance \emph{exploration} and \emph{exploitation}. Even though at first glance it may seem that each arm needs to be explored at least once, placing practical limits on large numbers of arms, there is a remedy to this as will be discussed shortly.

\textbf{The C$^2$UCB algorithm.}
Used to solve the contextual combinatorial bandit problem, the C$^2$UCB Algorithm~\cite{c2ucb} models the arms' scores as linearly dependent on their contexts: 
$r_{t}(i) = {\bm{\theta}}^\prime \bm{x}_t(i) + \varepsilon_t(i)$ for unknown zero-mean (subgaussian) random variable $\varepsilon_t$, unknown but fixed parameter $\bm{\theta}\in\mathbb{R}^d$, and known context $\bm{x}_t(i)$. It is crucial to notice that this implies that \textbf{all learned knowledge is contained in estimates of $\bm{\theta}$, which is shared between all arms, obviating the need to explore each arm}. Estimation of $\bm{\theta}$ can be achieved using ridge regression,\footnote{A standard linear regression with $L_2$ regularisation on the $\bm{\theta}$ coefficients, leading to well-posed optimisation. Equivalent to \emph{max a posteriori} Bayesian linear regression with a Gaussian prior.} with $|s_t|$ new data points $\{(\bm{x}_t(i), r_t(i))\}_{i \in s_t}$ available at round $t$, further \emph{accelerating the convergence rate} of the estimator $\hat{\bm{\theta}}$, over observing only one example as might be na\"ively assumed.

Point estimates on the expected scores can be made with $\bar{r}_t(i) = \hat{\bm{\theta}}_{t}^\prime \bm{x}_t(i)$,
where $\hat{\bm{\theta}}_{t}$ are trained coefficients of a ridge regression on arm $i$'s observed rewards against contexts. However, this quantity is oblivious to the variance in the score estimation. Intuitively, to balance out the exploration and exploitation, it is desirable to add an \emph{exploration boost} to the arms whose score we are less sure of (i.e., greater estimate variance). This suggests that the upper confidence bound (UCB) should be used, in place of the expected value, and which can be calculated~\cite{LinUCB} as:
\begin{eqnarray}
\hat{r}_t(i) &=&  \hat{\bm{\theta}}_{t}^\prime \bm{x}_t(i) + \alpha_t \sqrt{\bm{x}_t(i)^\prime\bm{V}_{t-1}^{-1}\bm{x}_t(i)}\enspace,\label{eq:linucb-value}
\end{eqnarray}
where $\alpha_t$ is the exploration boost factor, and $\bm{V}_{t-1}$ is the positive-definite $d\times d$ 
scatter matrix of contexts for the chosen arms up to and including round $t-1$. The first term of $\hat{r}_t(i)$ corresponds to arm $i$'s immediate reward, whereas its second term corresponds to its exploration boost, as its value is larger when the arm is sensitive to the context elements we are less confident of (i.e., the underexplored context dimension). Hence, by using $\hat{r}_t(i)$ in place of $\bar{r}_t(i)$, arms with contexts lying in the underexplored regions of context space are more likely to be chosen, as higher scores yield higher $g$, assuming that $g$ is monotonic increasing in the arm rewards.

Ideally, the super arm $s_t \in \mathcal{S}_t$ is chosen such that $g(\hat{\bm{r}}_t, \bm{X}_t, s_t)$ is maximised. However, it is sometimes computationally expensive to find such super arms. In such cases, it is often good enough to obtain a solution via some approximation algorithm where $g(\hat{\bm{r}}_t, \bm{X}_t, s_t)$ is \emph{near} maximum. With this criterion in mind, we now define an \emph{$\alpha$-approximation oracle}.

\begin{definition}
An \emph{$\alpha$-approximation oracle} is an algorithm $\mathcal{A}$ that outputs a super arm $\overline{s}=\mathcal{A}(\bm{r},\bm{X})$ with guarantee $g(\overline{s}, \bm{r},\bm{X})\geq \alpha \cdot \max_s g(s, \bm{r},\bm{X})$, for some $\alpha\in [0,1]$ and given input $\bm{r}$ and $\bm{X}$.  %
\end{definition}

Note that knapsack-constrained submodular programs are efficiently solved by the greedy algorithm (iteratively select a remaining cost-feasible arm with highest available score) with $\alpha = 1-1/e$.  C$^2$UCB is detailed in Algorithm~\ref{alg:c2ucb}.

\begin{algorithm}[t!]
\caption{The C$^2$UCB Algorithm}
\begin{algorithmic}[1]
\State Input: $\lambda, \alpha_1, \ldots, \alpha_T$
\State Initialize $\bm{V}_0 \gets \lambda \bm{I}_d$, $\bm{b}_0 \gets \bm{0}_d$
\For {$t \gets 1,\ldots,T$}
\State Observe $\mathcal{S}_t$
\State $\hat{\bm{\theta}}_{t} \gets \bm{V}_{t-1}^{-1}\bm{b}_{t-1}$ \Comment{estimate via ridge regression}
\For{$i \in [k]$}
\State Observe context $\bm{x}_t(i)$
\State $\hat{r}_t(i) \gets \hat{\bm{\theta}}_t^\prime \bm{x}_t(i)+ \alpha_t\sqrt{\bm{x}_t(i)^\prime\bm{V}_{t-1}^{-1}\bm{x}_t(i)}$
\EndFor
\State $s_t \gets \mathcal{A}(\hat{\bm{r}}_t, \bm{X}_t)$ \Comment{using $\alpha$-approximation oracle}
\State Play $s_t$ and observe $r_t(i)$ for all $i \in s_t$
\State $\bm{V}_{t} \gets \bm{V}_{t-1} + \sum_{i\in s_t}\bm{x}_t(i)\bm{x}_t(i)^\prime$ \Comment{regression update}
\State $\bm{b}_t \gets \bm{b}_{t-1} + \sum_{i \in s_t}r_t(i)\bm{x}_t(i)$ \Comment{regression update}
\EndFor
\end{algorithmic}
\label{alg:c2ucb}
\end{algorithm}

The performance of a bandit algorithm is usually measured by its \emph{cumulative regret}, defined as the total expected difference between the reward of the chosen super arm $\mathbb{E}[R_t(s_t)]$ and an optimal super arm $\max_{s \in \mathcal{S}_t}\mathbb{E}[R_t(s)]$ over $T$ rounds. However, such a metric is unfair to C$^2$UCB since its performance depends on the oracle's performance. This suggests to measure C$^2$UCB's performance with a new metric using the oracle's performance guarantee as its measuring stick, defined as follows.
\begin{definition}
Cumulative $\alpha$-regret is the sum of expected instantaneous regret,
$Reg_t^\alpha = \alpha\cdot \max_{s} g(s, \bm{r}^\star_t, \bm{X}_t) - g(\overline{s}_t, \bm{r}^\star_t, \bm{X}_t)$,
where $\overline{s}$ is a super arm returned by an $\alpha$-approximation oracle as a part of the bandit algorithm, while $\bm{r}^\star_t$ is a vector containing each arms' true expected scores. 
\end{definition}

When $g$ is assumed to be monotonic and Lipschitz continuous, \cite{c2ucb} claimed that C$^2$UCB enjoys $\tilde{O}(\sqrt{T})$ $\alpha$-regret.\footnote{The notation $\tilde{O}(\bm{\cdot})$ is equivalent to $O(\bm{\cdot})$ while ignoring logarithmic factors.} We have corrected an error in the original proof, as seen in our technical note~\cite{C2UCBProofNote}, confirming the $\tilde{O}(\sqrt{T})$ $\alpha$-regret. This expression is sub-linear in $T$, implying that the per-round average cumulative regret approaches zero after sufficiently many rounds. Consequently, online index selection based on C$^2$UCB comes endowed with a safety guarantee on worst-case performance: selections become at least as good as an $\alpha$-optimal policy (with perfect access to true  scores); and potentially much better than any fixed policy.

\section{MAB for Online Index Selection}
\label{sec:bandits}
Performant bandit learning for online index tuning demands arms covering important actions and no more, rewards that are observable and for which regret bounds are meaningful, and contexts and oracle that are efficiently computable and predictive of rewards. 
Each workload query is monitored for characteristics such as running time, query predicates, payload, etc. (Figure~\ref{fig:bandit_overall}). These observations feed into generation of relevant arms and their contexts. The learner selects a desired configuration which is materialised. After query return, the system identifies benefits of the materialised indices, which are then shaped into the reward signal for learning.

\textbf{Dynamic arms from workload predicates.} %
Instead of enumerating all column combinations, \emph{relevant} arms (indices) may be generated based on queries: combinations and permutations of query predicates (including join predicates), with and without inclusion of payload attributes from the selection clause. 
Such workload-based arm generation drastically reduces the action space,  %
and exploits natural skewness of real-life workloads that focus on small subsets of attributes over full tables~\cite{noDBSIGMOD2012, StratosExploration}. 
Workload-based arm generation is only viable due to dynamic arm addition (reflecting a dynamic action space) and is allowed by the bandit setting: we may define the set of feasible arms for each round at the start of the round.

\textbf{Context engineering.} %
Effective contexts are predictive of rewards, efficiently computable, and promote generalisation to previously unseen workloads and arms.
We form our context in two parts (see Figure~\ref{fig:bandit_overall}).

\textit{Context Part 1: Indexed column prefix.} We encode one context component per column. However unlike a bag-of-words or one-hot representation appropriate for text, similarity of arms depends on having similar column prefixes; common index columns is insufficient. This reflects a \textbf{\emph{novel bandit learning aspect of the problem.}} 
A context component has value $10^{-j}$ where $j$ is the corresponding column's position in the index, \emph{provided} that the column is included in the index and is a workload predicate column. The value is set to 0 otherwise, including if its presence only covers the payload. %

\begin{example}
Under the simplest workload (single query) in Figure~\ref{fig:bandit_overall}, our system generates six arms: four using different combinations and permutations of the predicates, two including the payload (covering indices).
Index IX5 includes column C1, but the context for C1 is valued as $0$, as this column is considered only due to the query payload. 
\end{example}

\textit{Context Part 2: Derived statistical information.} 
We represent statistical and derived information about the arms and workload, details available from the optimiser during query execution, and sufficient statistics for unbiased estimates.  %
This statistical information includes: a Boolean indicating a covering index, the estimated size of the index  divided by the database size (if not materialised already, 0 otherwise), and usage information of the index from previous rounds. This is shown in Figure~\ref{fig:bandit_overall} under D1, D2 and D3, respectively. 

\textbf{Reward shaping.} %
As the goal of physical design tuning tools is to %
minimise end-to-end workload time, we incorporate index creation time and query execution time into the reward for a workload. We omit index recommendation time, as it is independent of arm selection. However, we measure and report recommendation time of the MAB algorithm in our experiments. Recall that MAB depends only on observed execution statistics from implemented configurations and generalisation of the learned knowledge to unseen arms thereafter.

The implementation of the reward signal for an arm includes the execution time as a \emph{gain} $G_t(i, w_t, s_t)$ for a workload $w_t$ by each arm $i$ under configuration $s_t$.
By defining $\mathcal{U}(s, q)$ as the list of indices used by the query optimiser for query $q$ for a given configuration $s$, the gain by index $i$ for a query $q$ is defined by: 
\begin{eqnarray*}&&G_t(i, \{q\}, s_t) \\ &=& \left[C_{tab}(\tau(i), q, \emptyset) - C_{tab}(\tau(i), q, \{i\})\right]\mathbbm{1}_{\mathcal{U}(s,q)}(i)\,, \end{eqnarray*} 
where $\tau(i)$ is the table which $i$ belongs to and $C_{tab}(\tau(i), q, \emptyset)$ represents the full table scan time for table $\tau(i)$ and query $q$.\footnote{Due to the reactive nature of multi-armed bandits, we mostly observe a full table scan time for each table $\tau(i)$ and query $q$. When we do not observe this, we estimate it with the maximum secondary index scan/seek time.} The gain for a workload is related to the gain for individual query by:
$$ G_t(i, w_t, s_t) = \sum_{q\in w_t}G_t(i, \{q\}, s_t)\, .$$
By this definition, 
gain $G_t(i,w_t, s_t)$ will be $0$ if $i$ is not used by the optimiser in the current round $t$ and can be negative if the index creation leads to a performance regression. Creation time of $i$ is taken as a negative reward, only if $i$ is materialised in round $t$, and is $0$ otherwise:
$$r_{t}(i) = G_t(i,w_t, s_t) - C_{cre}(s_{t-1}, \{i\})\enspace.$$
Minimising the end-to-end workload time, or rather, maximising the end-to-end workload time gained, is the goal of the bandit. As defined earlier, the total workload time $C_{tot}$ is the sum of \emph{execution}, \emph{recommendation} and \emph{creation} times accumulated over rounds. As such, minimising each round's summand is an equivalent problem. Modifying the execution time to the time gain while ignoring the recommendation time yields per-round super arm reward of:
\begin{align*}
    R_t(s_t) &= C_{exc}(w_t, \emptyset) - [C_{exc}(w_t,s_t) + C_{cre}(s_{t-1},s_t)]\\
    &\approx \sum_{i\in s_t} G_t(i,w_t,s_t) - \sum_{i \in s_t}C_{cre}(s_{t-1}, \{i\})\\
    &= \sum_{i\in s_t}r_t(i)\enspace.
\end{align*}
Selection of the execution plan depends on the query optimiser, and as noted, %
the query optimiser may resolve to a sub-optimal query plan. As we show, the bandit is nonetheless resilient as it can quickly recover from any such performance regressions. %
Observed execution times encapsulate real-world effects \eg the interaction between queries, application properties, run-time parameters, etc. 
Since the end-to-end workload time includes the index creation and query execution times, we are indirectly optimising for both efficiency and the quality of recommendations.

\textbf{A greedy oracle for super-arm selection.} Recall that C$^2$UCB leverages a near-optimal oracle to select a super arm, based on individual arm scores%
~\cite{c2ucb}. As a sum of individual arm rewards, our super-arm reward has a (sub)modular objective function and (as easily proven) exhibits monotonicity and Lipschitz continuity. Approximate solutions to maximise submodular (diminishing returns) objective functions can be obtained with greedy oracles that are efficient and near-optimal~\cite{nemhauser1978analysis}. Our implementation uses such an oracle combined with filtering to encourage diversity. Initially, arms with negative scores are pruned. Then arm selection and filtering steps alternate, until the memory budget is reached. In the selection step, an arm is selected greedily based on individual scores. The filtering step filters out arms that are no longer viable under the remaining memory budget, or those that are already covered by the selected arms based on prefix matching. If a covering index is selected for a query, all other arms generated for that query will be filtered out. %
Note that filtering is a temporary process that only impacts the current round.

\begin{algorithm}[t!]
\caption{MAB Simulation for Index Tuning}\label{euclid}
\begin{algorithmic}[1]
\State $\textbf{QS} \gets QueryStore()$ \Comment{keeps query information}
\While{(TRUE)}
\State $\textbf{queries} \gets getLastRoundWorkload()$ 
\ForAll{$queries$}
\If{(isNewTemplate)} 
\State $QS.add(query)$
\Else
\State $QS.update(query)$
\EndIf
\EndFor
\State $\textbf{QoI} \gets QS.getQoI()$ \Comment{get queries of interest}
\State $\textbf{arms} \gets generateArms(QoI)$
\State $\textbf{contexts} \gets generateContext(arms, QoI)$
\State $\textbf{s\textsubscript{t}} \gets \textbf{C\textsubscript{2}UCB}.recommend(arms, contexts)$
\State $\textbf{C\textsubscript{cre}} \gets materialise(s\textsubscript{t})$
\State $\textbf{C\textsubscript{exc}} \gets executeCurrentWorkload()$ 
\State $\textbf{C\textsubscript{2}UCB}.updateReward(C\textsubscript{cre}, C\textsubscript{exc})$ 
\EndWhile
\end{algorithmic}
\label{algorithm}
\end{algorithm}

\textbf{Bandit learning algorithm.} 
Algorithm~\ref{algorithm} shows the MAB algorithm
 which summarises workload information using templates; these track frequency, average selectivity, first seen and last seen times of the queries which help to generate the best set of arms per round (\ie QoI). The context is updated after each round based on the workload and selected set of arms. The bandit then selects a set of arms for this round.
 The set of arms chosen form a configuration to be materialised within the database.
 Once the new configuration is in place, a new set of queries will be executed. %
 To support shifting workloads, where users' interests change over time, the learner can forget learned knowledge depending on the workload shift intensity (\ie number of newly introduced query templates). 

\section{Experimental Evaluation}
\label{sec:evaluation}
We evaluate our MAB framework across a range of widely used industrial benchmarks, comparing it to a state-of-the-art physical design tool shipped with a commercial database product referred to as the Physical Design Tool (PDTool). This is a mature product, proven to outperform other physical design tools available on the market.

\begin{figure*}[t]
\centering
\begin{minipage}{0.19\textwidth}
\centering\includegraphics[width=\textwidth]{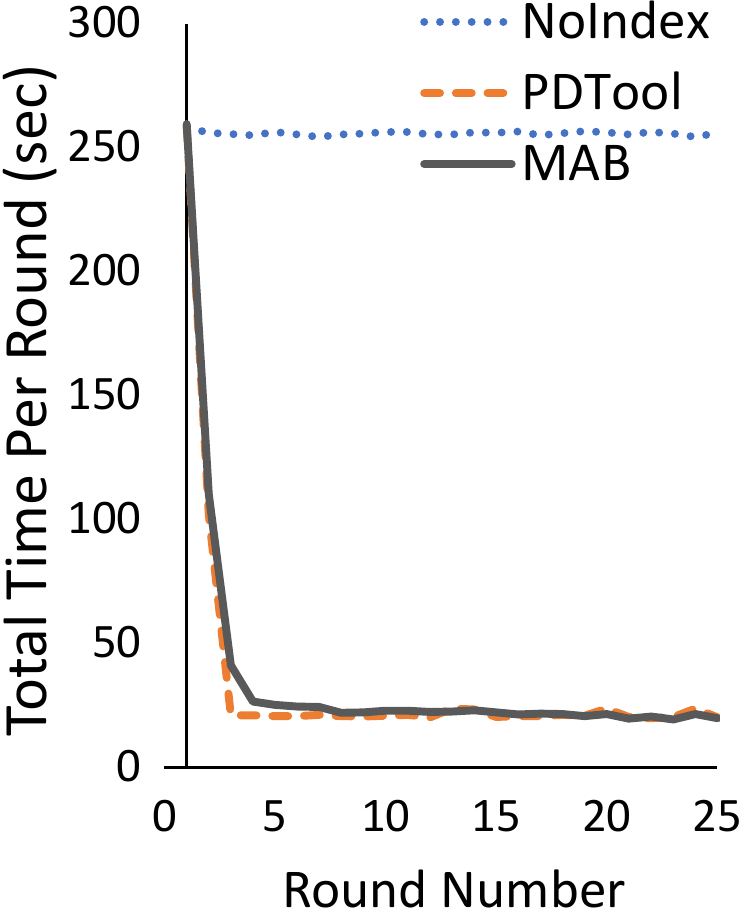}\\ \textbf{(a)}
\end{minipage}\hfill
\begin{minipage}{0.19\textwidth}
\centering\includegraphics[width=\textwidth]{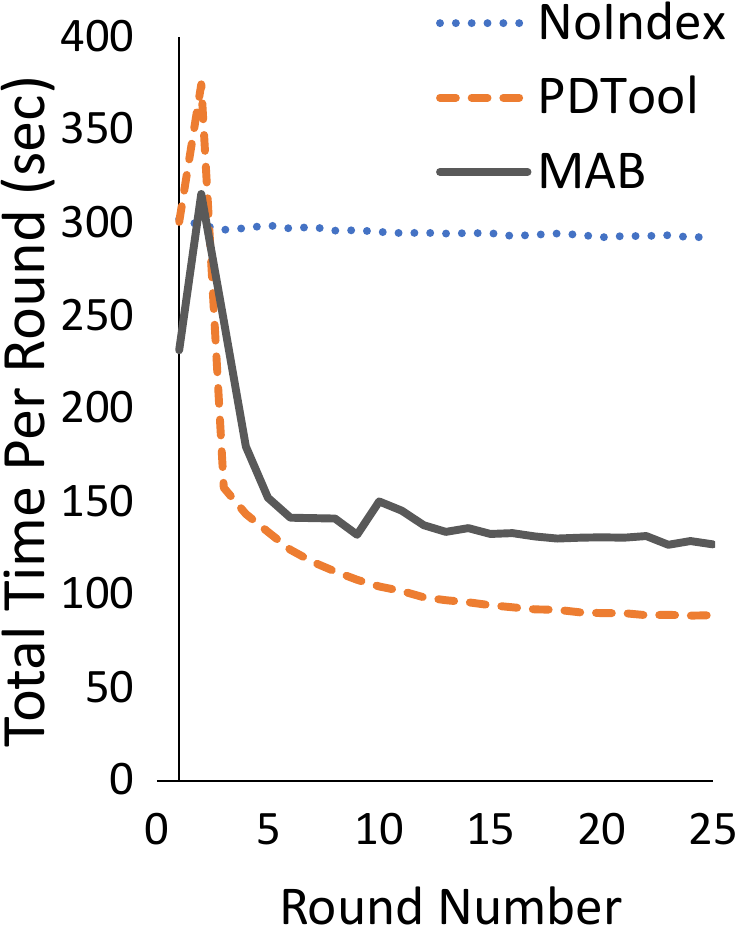}\\ \textbf{(b)}
\end{minipage}\hfill
\begin{minipage}{0.19\textwidth}
\centering\includegraphics[width=\textwidth]{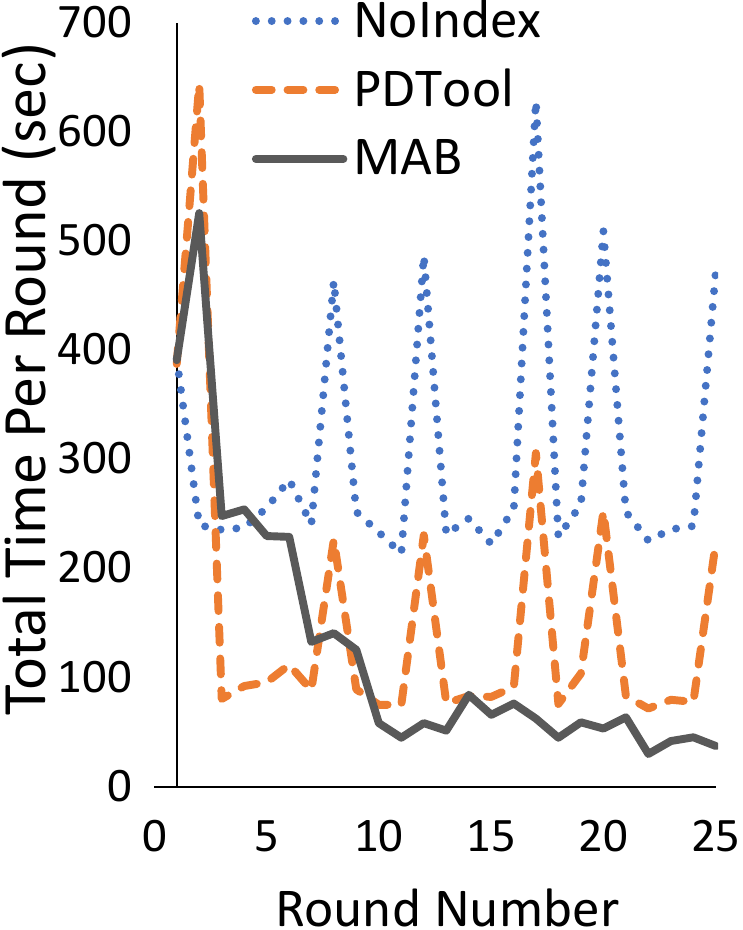}\\ \textbf{(c)}
\end{minipage}\hfill
\begin{minipage}{0.19\textwidth}
\centering\includegraphics[width=\textwidth]{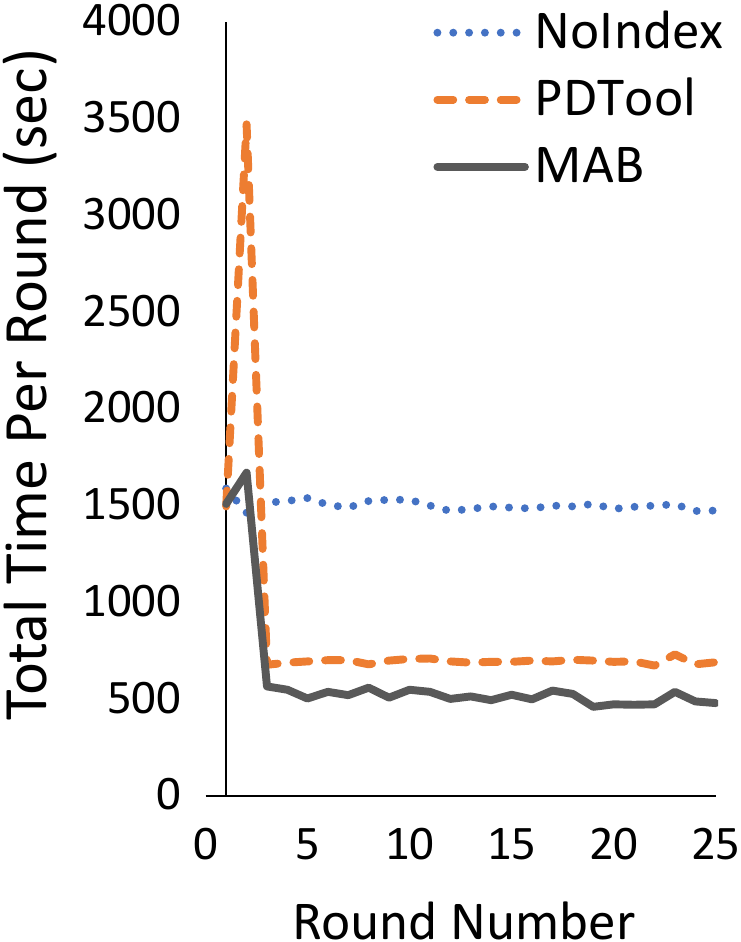}\\ \textbf{(d)}
\end{minipage}\hfill
\begin{minipage}{0.19\textwidth}
\centering\includegraphics[width=\textwidth]{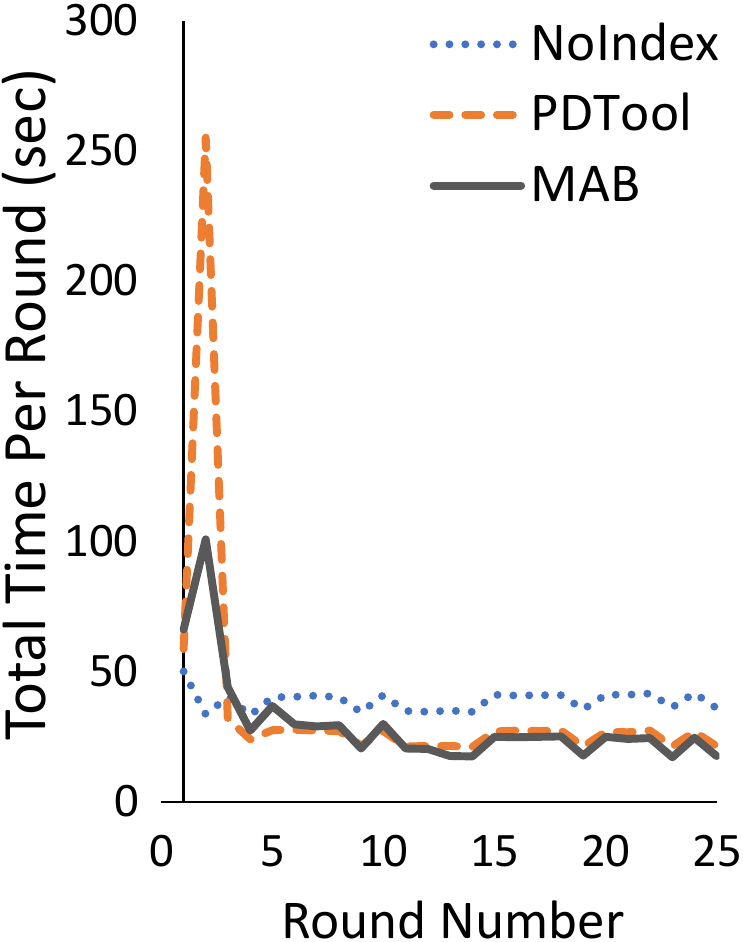}\\ \textbf{(e)}
\end{minipage}\\
\caption{MAB vs. PDTool Convergence for \emph{static} workloads: (a) SSB, (b) TPC-H, (c) TPC-H Skew, (d) TPC-DS and (e) IMDb.}
\label{fig:static-convergence}
\end{figure*}

\begin{figure}[t]
\centering
\includegraphics[width=\columnwidth]{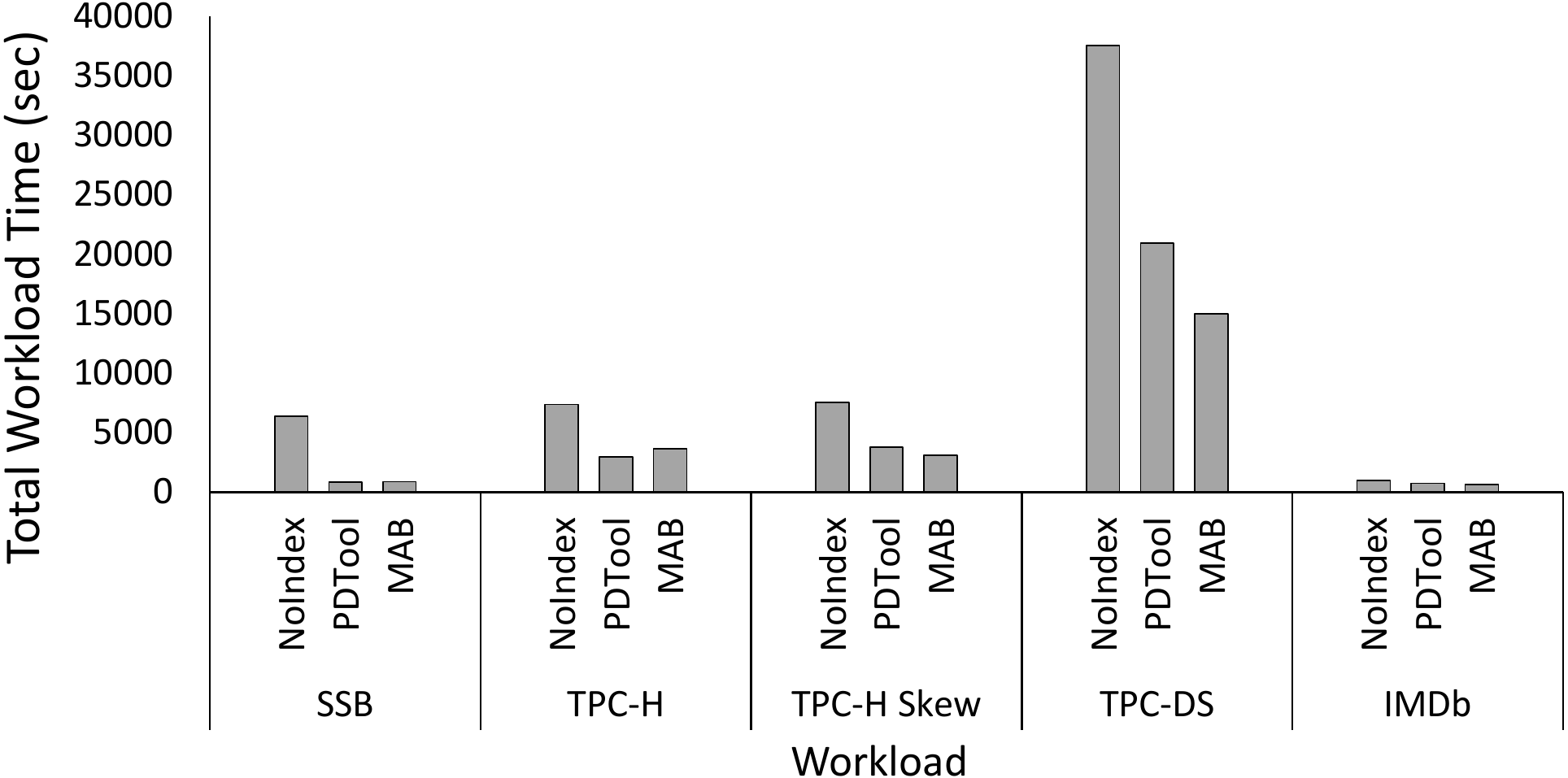}\\
\caption{MAB vs. PDTool total end-to-end workload time for \emph{static} workloads.}
\label{fig:static-total}
\end{figure}

\subsection{Experimental Setup}
\textbf{Benchmarks.}
We use five publicly available benchmarks: TPC-H (with uniform distribution)~\cite{tpch} and TPC-H Skew~\cite{tpchskew} with zipfian factor 4, allowing the reader to understand the impact of data skewness when all the other aspects are kept identical; TPC-DS~\cite{tpcds}, which demonstrates the solution fitness under a large number of candidate configurations; SSB~\cite{SSB} with easily achievable high index benefits; and finally, Join Order Benchmark (JOB) with IMDb dataset (a real-world dataset)~\cite{HowGoodQO} (henceforth referred to as IMDb) a challenging workload for index recommendations, with index overuse leading to performance regressions.

Unless stated otherwise, all experiments use scale factor (SF) 10, resulting in approximately 10GB of data per workload, except in the case of the IMDb dataset which has a fixed size of 6GB. We consider three broad types of workloads, allowing us to compare different aspects of the recommendation process:

\emph{Static}: The workload sequence is known in advance, and repeating over time (modelling workloads used for reporting purposes). In absence of dynamic environment complexities, this simpler setting allows us to single-out the effectiveness (ability to find a better configuration) and the efficiency (additional overhead) of the MAB search strategy.
    
\emph{Dynamic shifting}: The region of interest shifts over time from one group of queries to another (modelling data exploration). This experiment evaluates the adaptation speed to workload shifts and the cost of exploration when adapting.

\emph{Dynamic random}: A query sequence is chosen entirely at random (modelling more dynamic settings, such as cloud services). Dynamic random experiments test the delicate balance between swift and careful adaptation under returning workloads, which can lead to unwanted index oscillations.

Both PDTool and MAB are given a memory budget approximately equal to the size of the data (1x; 10GB for SF 10 datasets and 6GB for IMDb dataset) for the creation of secondary indices. We have experimented with different memory budgets ranging from 0.25x to 2x (since benefits of additional memory seem to diminish beyond a 2x limit) under TPC-H and TPC-H skew benchmarks, and observed the same patterns throughout that range. We have naturally picked the middle of the active region (1x) as our memory budget. All these workloads come with original primary and foreign keys that influence the choice of indices. We grant the aforementioned memory budget on top of this. 

In search of the best possible design, we do not constrain the running time of PDTool, with one exception: In TPC-DS dynamic random, PDTool was uncompetitive due to long running times,\footnote{A single PDTool invocation took around 8 hours (default limit). The total recommendation time was around 40 hours, which is not competitive compared to the end-to-end workload time of 4 hours under MAB.} hence the PDTool running time of each invocation was restricted to 1 hour. All proposed indices are materialised and queries invoked over the same commercial DBMS in both cases (MAB and PDTool). 

Across experiments, each group of templatized queries is invoked over rounds, producing different query instances. For static and dynamic settings, PDTool is invoked every time after the first round of new queries, with those queries given as the training workload, since this workload will become representative of future rounds. This setting is somewhat unrealistic and favourable for PDTool, since in real-life the PDTool will seldom truly have knowledge of the representative workload (\ie what is yet to arrive in the future), advantaging the PDTool in our experiments. However, it presents a viable comparison against the workload-oblivious MAB. Bandits, do not use any workload information ahead of time but instead observe workload sequence and react accordingly. 

\textbf{Hardware.}
All experiments are performed on a server equipped with 2x 24 Core Xeon Platinum 8260 at 2.4GHz, 1.1TB RAM, and 50TB disk (10K RPM) running Windows Server 2016. %
We report cold runs, clearing database buffer caches prior to every query execution. 

\textbf{Metrics.}
In addition to reporting total end-to-end workload time for all rounds, we also report the total workload time per round used to demonstrate the convergence of different tools. Additionally, we present the total workload time broken down by recommendation time (when invoking the PDTool or the MAB framework), index creation time, and workload execution time. For completeness, we show original query times, without any secondary indices (denoted as NoIndex).
In addition to convergence graphs of individual benchmarks, we present a summary graph with total end-to-end workload time for all rounds under MAB and PDTool tuning of SSB, TPC-H (uniform), TPC-H skew, TPC-DS and IMDb benchmarks.

\subsection{Experimental Results}
\subsubsection{MAB versus the existing physical design tool}
We report on wide ranging empirical comparisons of MAB and PDTool.\\[-0.8em]

\begin{figure*}[t]
\centering
\begin{minipage}{0.19\textwidth}
\centering\includegraphics[width=\textwidth]{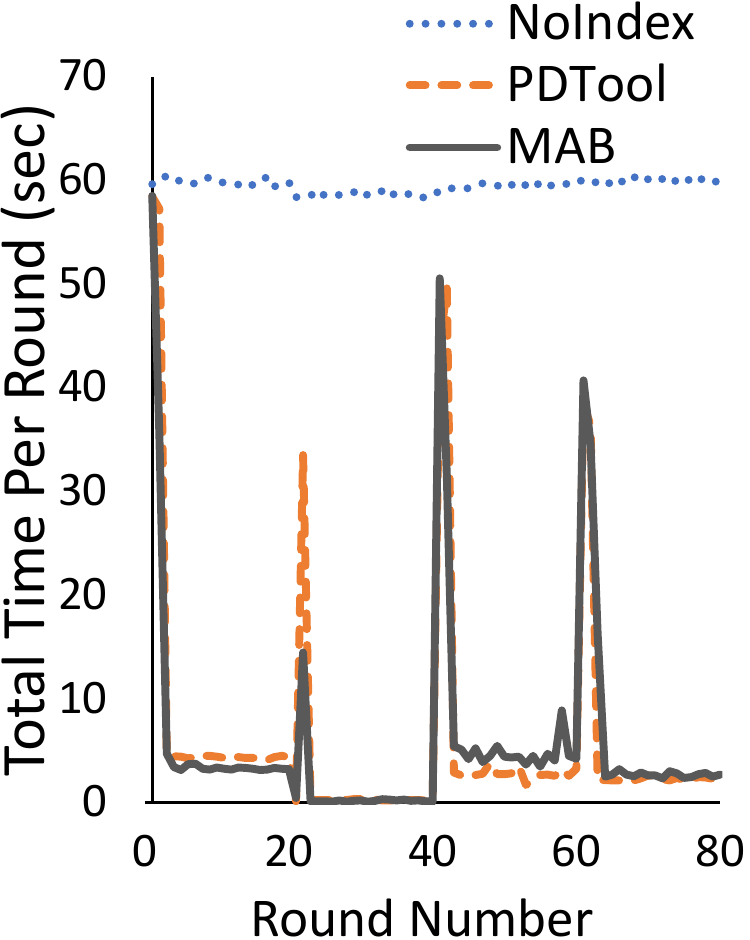}\\ \textbf{(a)}
\end{minipage}\hfill
\begin{minipage}{0.19\textwidth}
\centering\includegraphics[width=\textwidth]{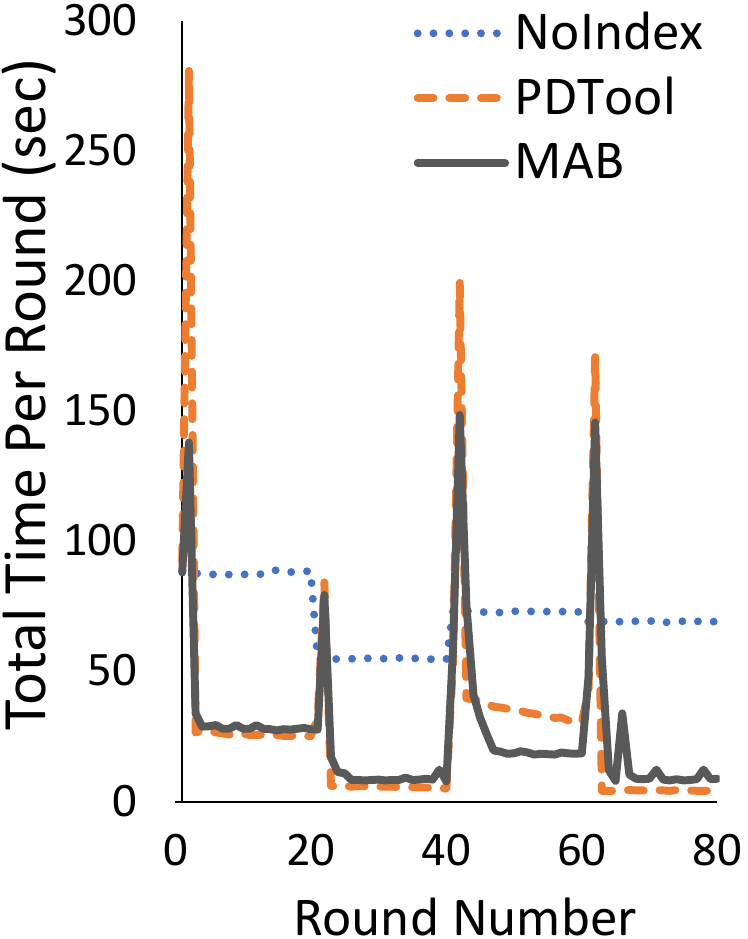}\\ \textbf{(b)}
\end{minipage}\hfill
\begin{minipage}{0.19\textwidth}
\centering\includegraphics[width=\textwidth]{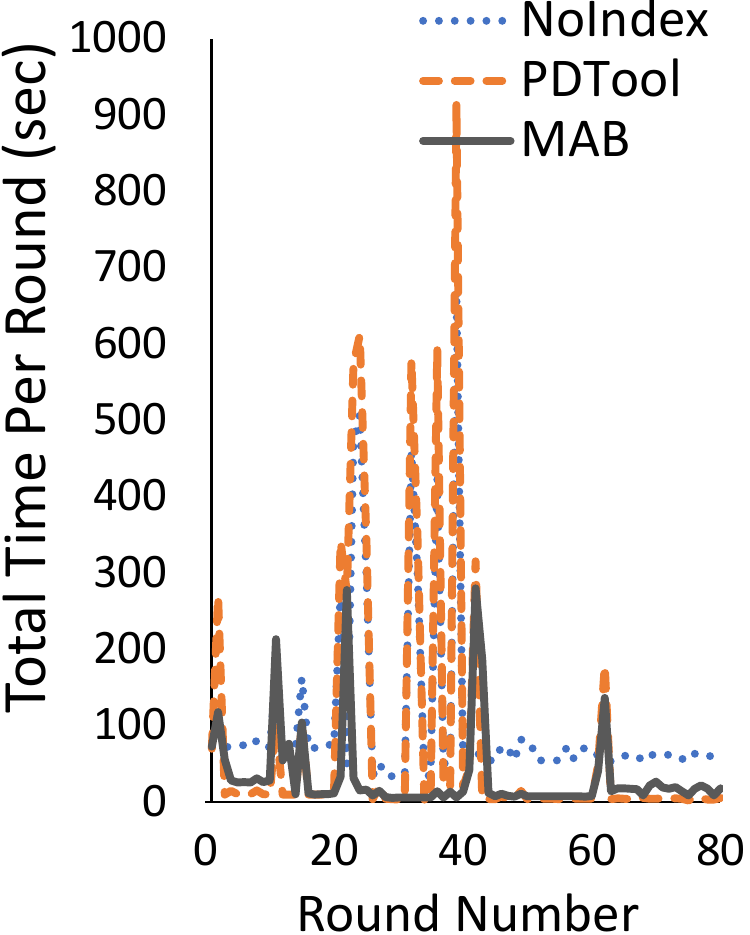}\\ \textbf{(c)}
\end{minipage}\hfill
\begin{minipage}{0.19\textwidth}
\centering\includegraphics[width=\textwidth]{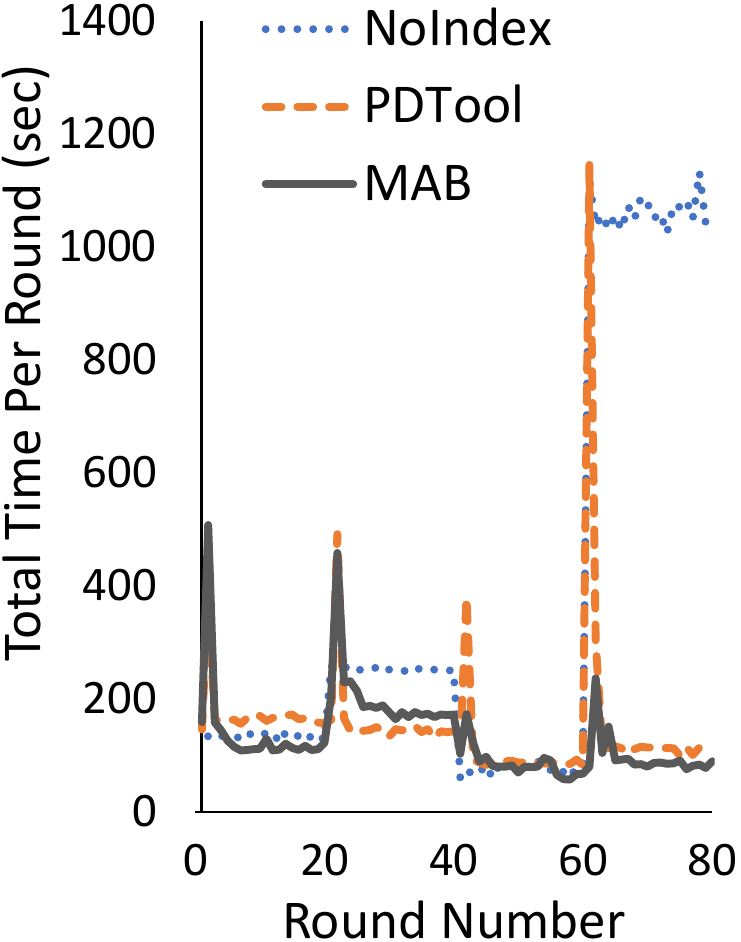}\\ \textbf{(d)}
\end{minipage}\hfill
\begin{minipage}{0.19\textwidth}
\centering\includegraphics[width=\textwidth]{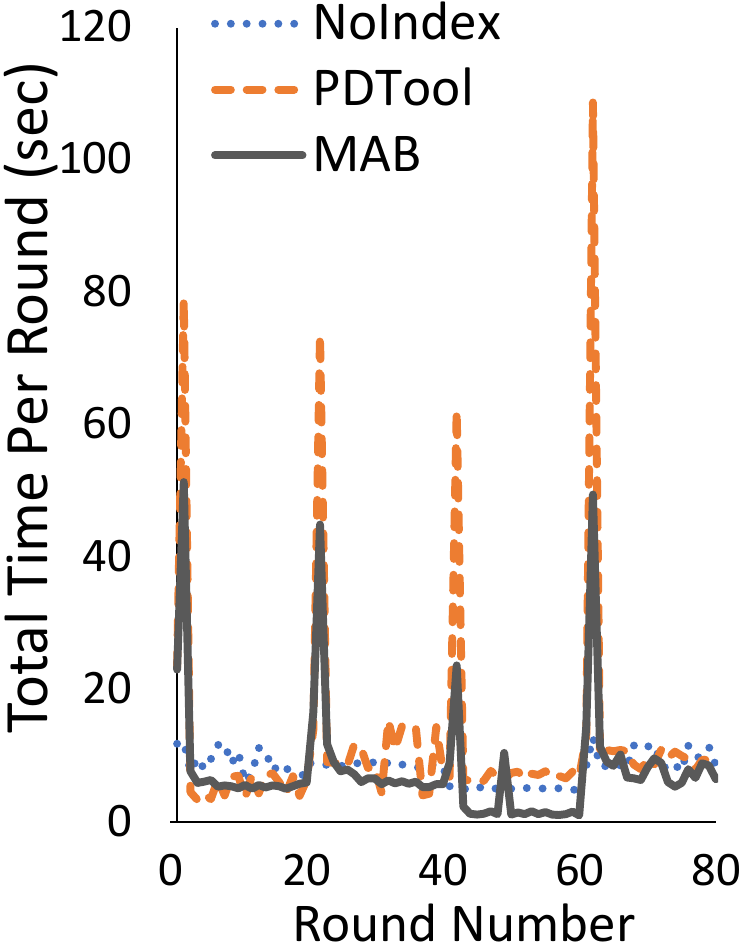}\\ \textbf{(e)}
\end{minipage}\\
\caption{MAB vs. PDTool Convergence for \emph{dynamic shifting} workloads: (a) SSB, (b) TPC-H, (c) TPC-H Skew, (d) TPC-DS and (e) IMDb.}
\label{fig:dynamic-convergence}
\end{figure*}

\begin{figure}[t]
\centering
\includegraphics[width=\columnwidth]{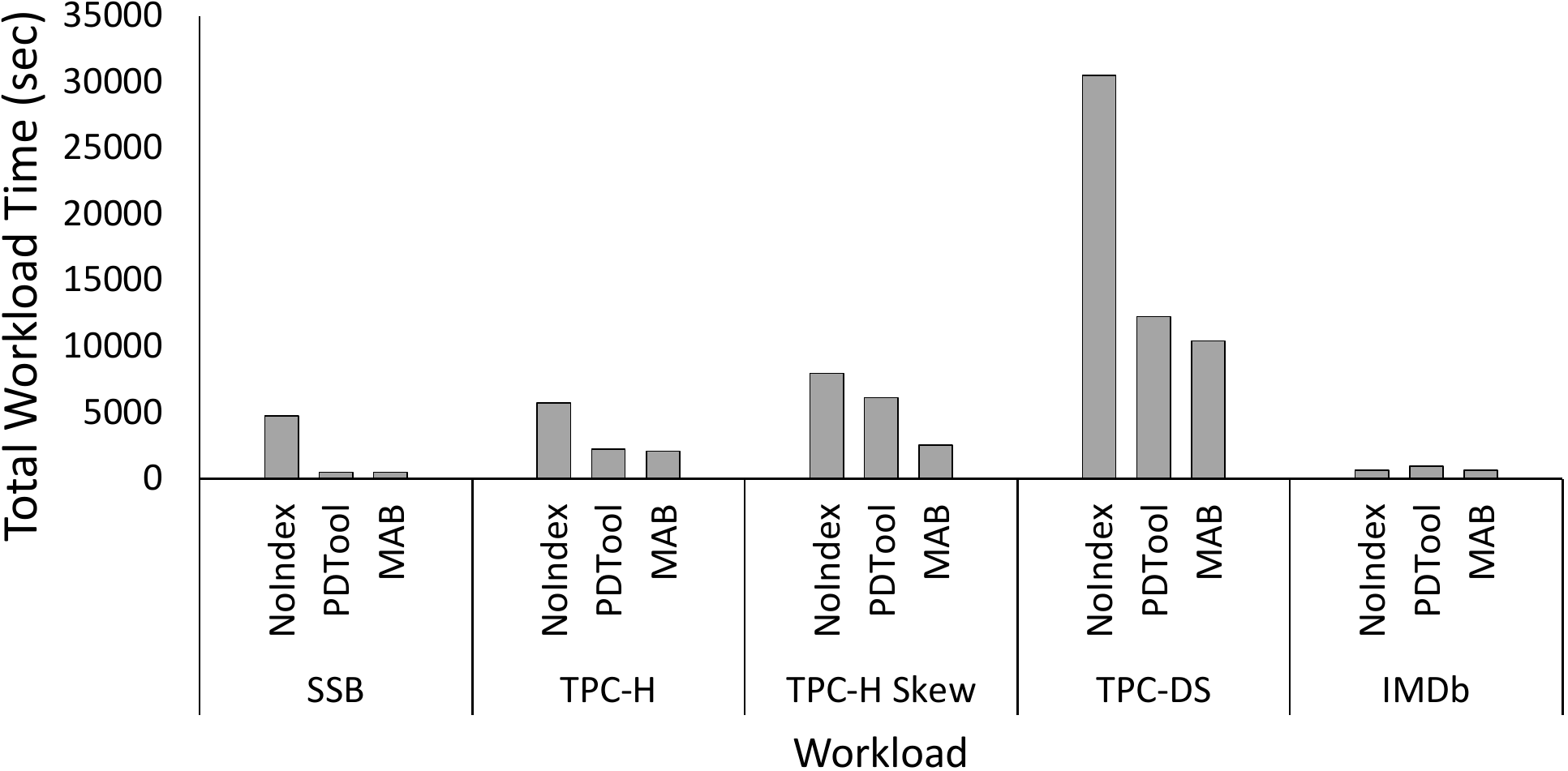}
\caption{MAB vs. PDTool total end-to-end workload time for \emph{dynamic shifting} workloads.}
\label{fig:dynamic-total}
\end{figure}

\textbf{Static workloads.} Static workloads over uniform datasets are the best case for offline physical design tools, as a pre-determined workload sequence may perfectly represent future queries. However, when underlining data is skewed, recommendations based on a pre-determined workload alone can have unfavourable outcomes. While used for reporting, static workloads do not reflect modern dynamic workloads (\eg data exploration). In static workloads, 
all query templates in the benchmark (22, 13, 99 and 33 templates for TPC-H, SSB, TPC-DS and IMDb, respectively) are invoked once every round, each with a different query instance of the template, for a total of 25 rounds, providing sufficient time to observe  convergence.

Figure~\ref{fig:static-total} displays overall workload time (including recommendation and index creation time) for all 25 rounds under MAB and PDTool. For skewed datasets (TPC-H Skew, TPC-DS, IMDb) MAB outperforms PDTool. MAB shows over 17\%, 28\% and 11\% performance gain against PDTool, under TPC-H Skew, TPC-DS, IMDb benchmarks, respectively. Under uniform datasets (TPC-H and SSB), both MAB and PDTool provide significant performance gains over NoIndex (over 50\% and 85\%, respectively), while PDTool outperforms the MAB (by 19\% and 5\%). This is not surprising since for uniform, static experiments usually align with PDTool assumptions and the future can be perfectly represented by a pre-determined workload.

Convergence plots in Figure~\ref{fig:static-convergence}(a--e), show MAB's gradual improvement over 25 rounds. Both MAB and PDTool have large spikes after the first round for all the workloads (spikes are less visible in SSB due to the small number of simple indices created). For both tools, this is due to recommendation and creation of indices. However, MAB might drop proposed indices and create new ones later on, generating relatively smaller spikes in subsequent rounds. Nonetheless, MAB efficiently balances the exploration of new indices, reducing exploration with time.

\emph{What is the best search strategy?} 
Comparison of execution times in the final round of the static experiment provides a clear idea about the benefit of using execution cost guided search. As evident from Figure~\ref{fig:static-convergence}(a--e), in 4 out of 5 cases, MAB converges to a better configuration than PDTool. MAB provides over 5\%, 84\%, 31\% and 19\% better execution time by the last round (25\textsuperscript{th}) compared to PDTool under SSB, TPC-H Skew, TPC-DS and IMDb, respectively.

For TPC-H skew, PDTool misses an index on $Orders.O\_custkey$. This index boosts the performance of some queries (Q22 in particular) which MAB correctly detects and materialises. Missing this leads to large execution times in a few rounds including the last round (8, 12, 17, 20 and 25) for PDTool. These experiments illustrate the risk of relying on the query optimiser and imperfect statistics as a single source of truth. %

The only case when MAB is outperformed by the PDTool is under TPC-H (PDTool delivers over 21\% better execution time by the last round): different indices are proposed, as our current MAB framework does not support an index merging phase employed by some physical design tools~\cite{chaudhuri1999merging}. Instead, MAB uses individual queries to propose index candidates. We plan to address index merging in future work. \\[-0.8em]

\begin{figure*}[t]
\centering
\begin{minipage}{0.19\textwidth}
\centering\includegraphics[width=\textwidth]{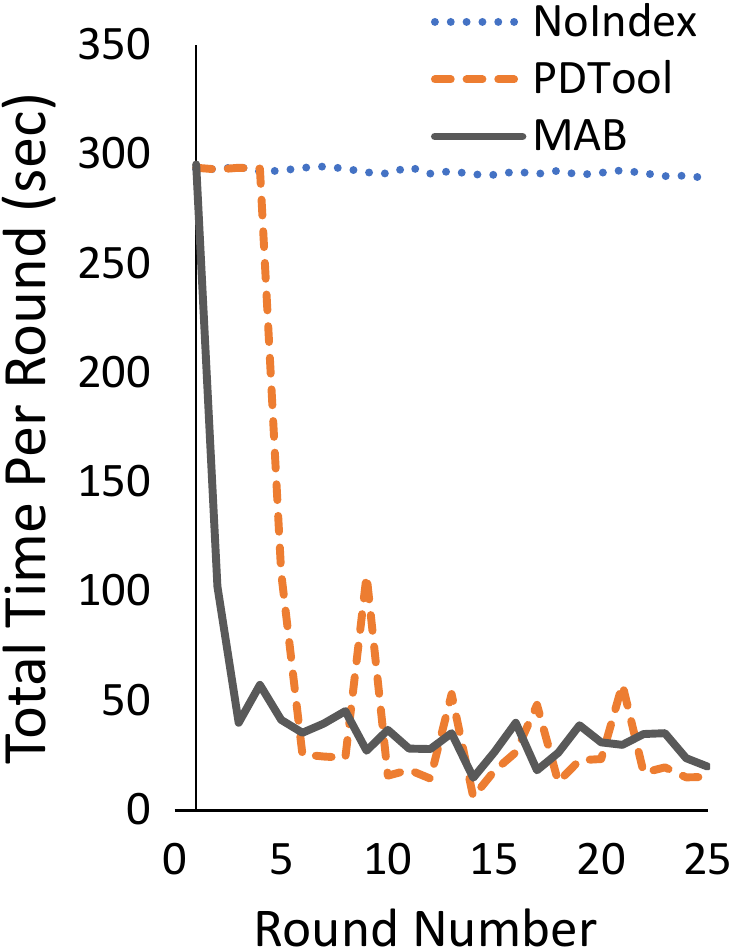}\\ \textbf{(a)}
\end{minipage}\hfill
\begin{minipage}{0.19\textwidth}
\centering\includegraphics[width=\textwidth]{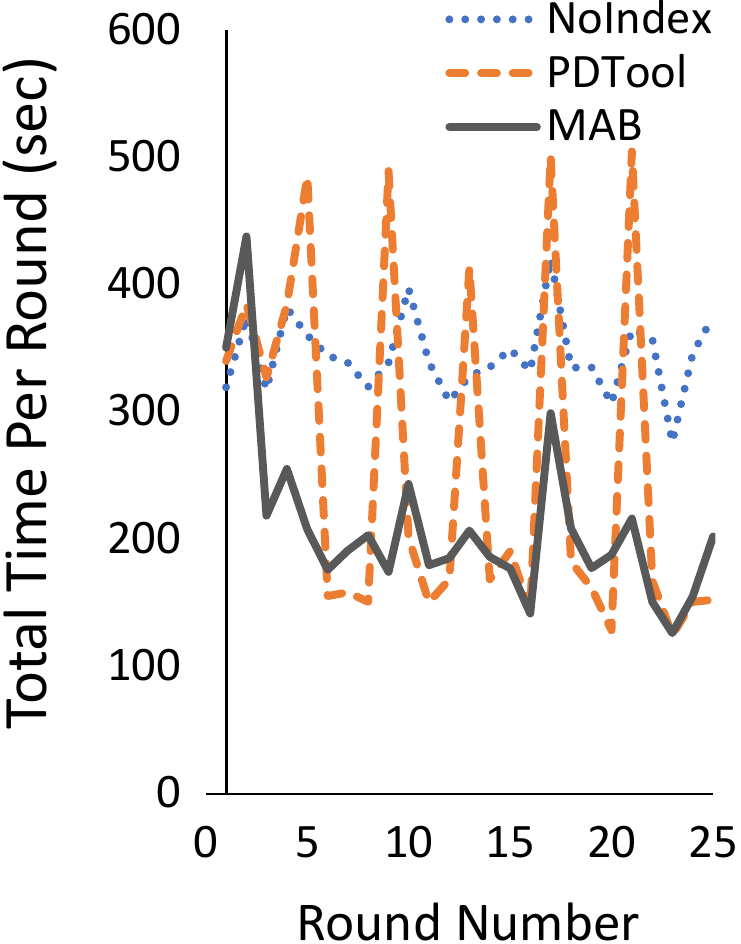}\\ \textbf{(b)}
\end{minipage}\hfill
\begin{minipage}{0.19\textwidth}
\centering\includegraphics[width=\textwidth]{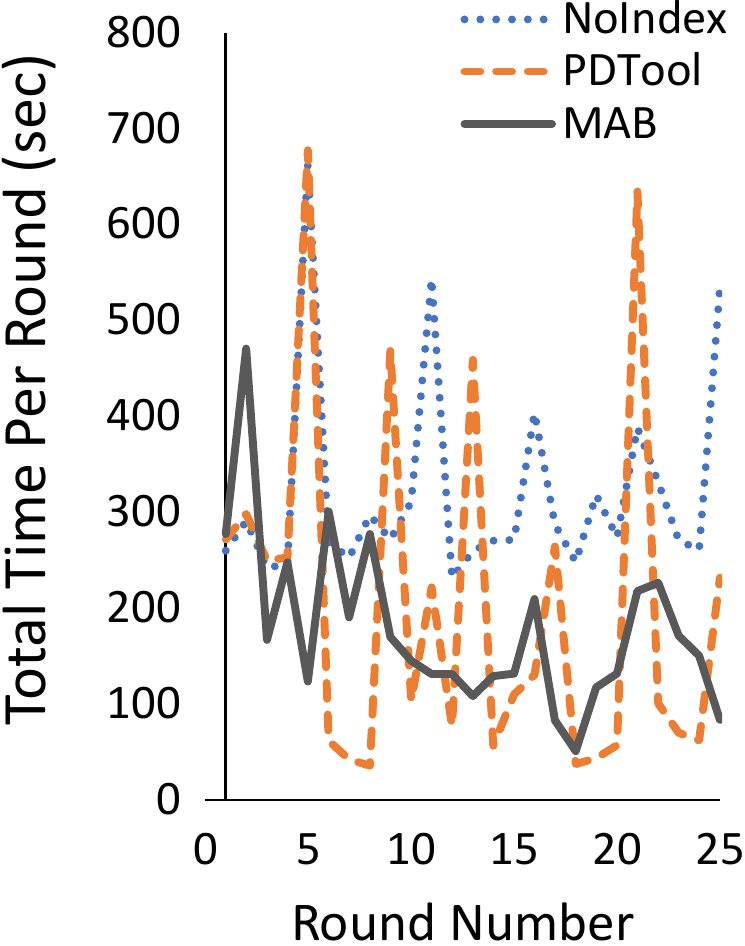}\\ \textbf{(c)}
\end{minipage}\hfill
\begin{minipage}{0.19\textwidth}
\centering\includegraphics[width=\textwidth]{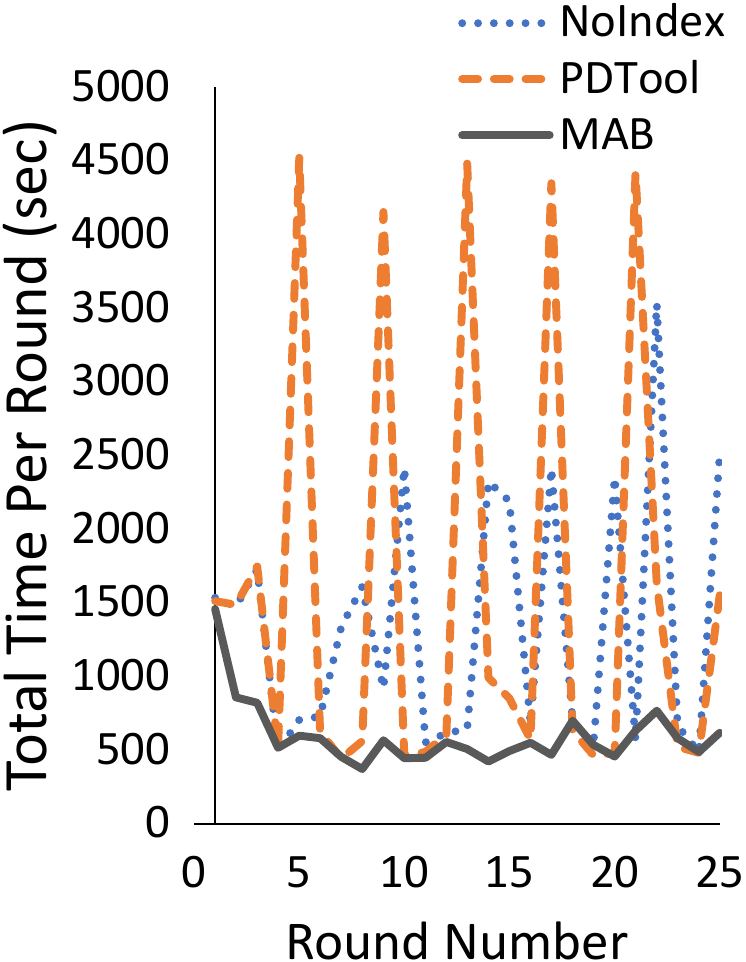}\\ \textbf{(d)}
\end{minipage}\hfill
\begin{minipage}{0.19\textwidth}
\centering\includegraphics[width=\textwidth]{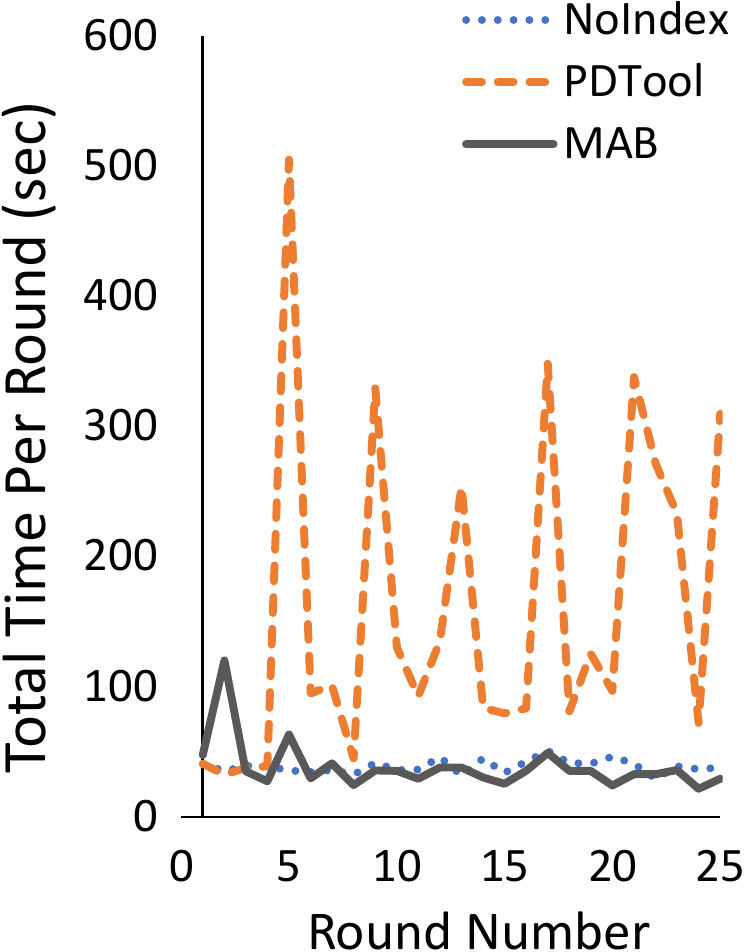}\\ \textbf{(e)}
\end{minipage}\\
\caption{MAB vs. PDTool Convergence for \emph{dynamic random} workloads: (a) SSB, (b) TPC-H, (c) TPC-H Skew, (d) TPC-DS and (e) IMDb}
\label{fig:random-convergence}
\end{figure*}

\begin{figure}[t]
\centering
\includegraphics[width=\columnwidth]{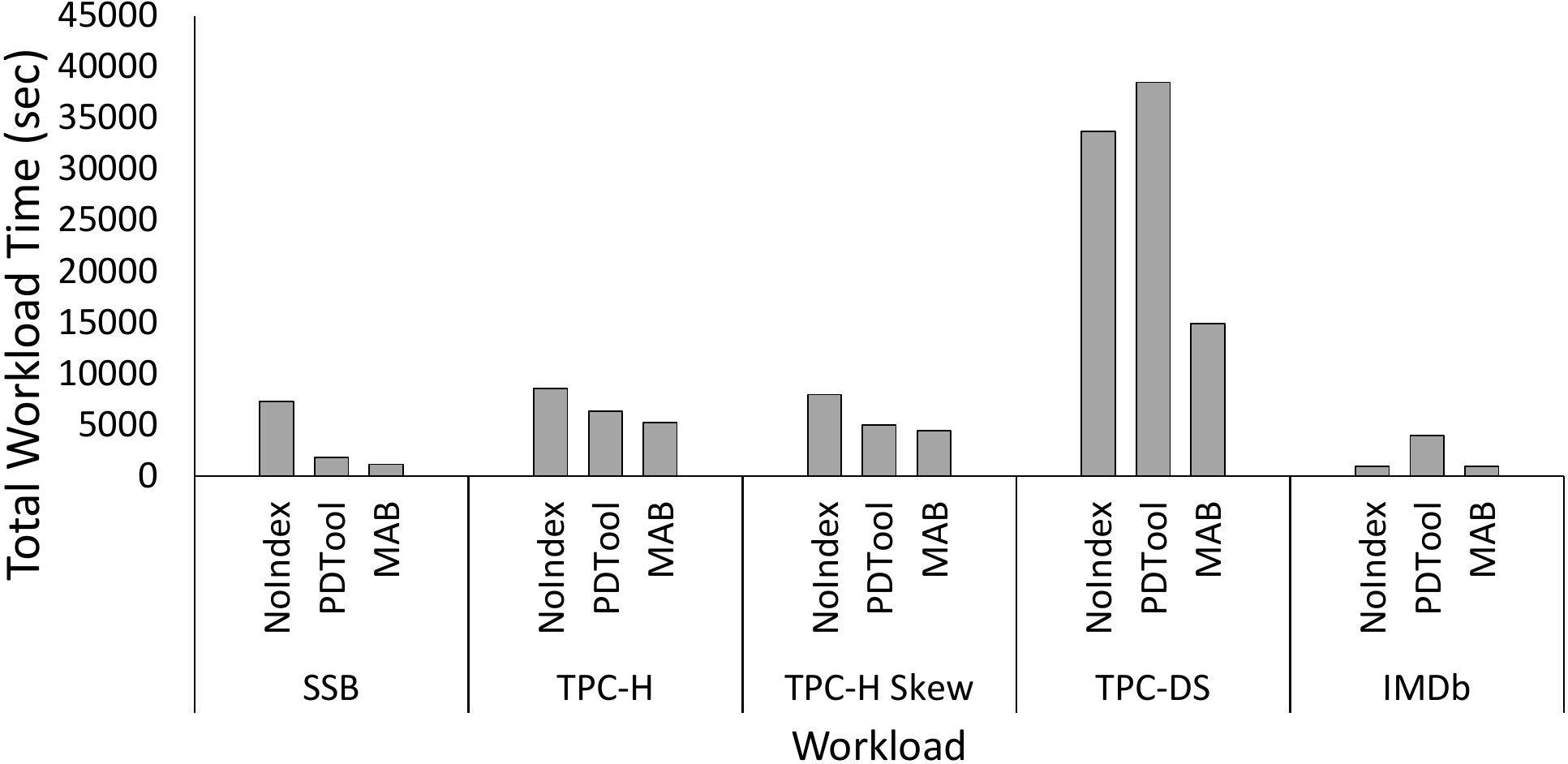}\\ 
\caption{MAB vs. PDTool total end-to-end workload time for \emph{dynamic random} workloads.}
\label{fig:random-total}
\end{figure}

\textbf{Dynamic shifting workloads.} Under the dynamic shifting workloads, all query templates in the benchmark are randomly divided into 4 equal-sized groups.
A group of query templates is then executed for 20 rounds, after which the workload switches to a new group of unseen queries (no overlap with the previous queries). %
When the workload switches, PDTool is invoked and trained on the new sequence of queries (whose templates will be used in the next 19 rounds).\footnote{Again, this relaxation assumes a DBA with knowledge that the previous workload will not be repeated, placing PDTool at an advantage. In reality, proposing training workloads might be much more challenging for dynamic workloads.} Thus, PDTool is invoked four times in total (in rounds 2, 22, 42, 62). On the other hand, the MAB framework does not assume any workload knowledge.

Figure~\ref{fig:dynamic-total} displays MAB's end-to-end workload time as substantially lower compared to the alternatives, under all benchmarks. MAB provides over 3\%, 6\%, 58\%, 14\% and 34\% speed-up compared to PDTool, under SSB, TPC-H, TPC-H Skew, TPC-DS and IMDb, respectively.

Interestingly, NoIndex performs better than PDTool against the IMDb workload. PDTool has a higher total workload time as well as higher execution time compared to NoIndex. NoIndex provides 3.5\% (24 seconds) speedup in execution time over PDTool. This is mainly due to misestimates of the optimiser~\cite{das2019automatically}. As an example (out of many), query 18 takes less than 1 second under NoIndex, whereas with the created indices by PDTool some instances of this query take around 7-8 seconds due to a suboptimal plan chosen favoring the index usage.
This affects both MAB and PDTool, but MAB identifies the indices with a negative impact based on the reward and drops them. For the IMDb workload which does not get much support from indices, MAB provides 3\% total performance gain and 26\% execution time gain compared to NoIndex.

One can easily observe the workload shifts in Figure~\ref{fig:dynamic-convergence}(a--e) due to the spikes in rounds 2, 22, 42, and 62. For PDTool, this is due to the invocation of PDTool and index creation after the workload shifts. Similar spikes can be seen in the MAB line with automatic detection of workload shifts. Further random spikes can be observed, for PDTool, from rounds 20-40 in TPC-H skew and rounds 30-40 under IMDb, due to the issues discussed in the previous paragraphs (Q22 in TPC-H Skew, Q18 in IMDb).\\[-0.8em]

\textbf{Dynamic random workloads.} We simulate modern data analytics workloads that are truly ad-hoc in nature. For instance, cloud providers, hosting millions of databases, seldom can detect representative queries, since they frequently change~\cite{das2019automatically}. In such cases, it is common to invoke the PDTool periodically (\eg nightly or weekly) using queries since the last invocation as the training workload. In this setting, we invoke the PDTool every 4 rounds, using queries from the last 4 rounds as the representative workload. %
In the dynamic random setting, the number of total training queries in the complete sequence is similar to the number of queries we had in the static setting. However, we have no control over the selection of queries for the workload and they are chosen completely randomly. The sequence is then divided into 25 equal-sized rounds. In all cases, the round-to-round repeat workload was between 45-54\%.

As shown in Figure~\ref{fig:random-total}, again we see a considerably lower total workload time of MAB compared to PDTool. MAB provides over 37\%, 17\%, 11\%, 61\% and 75\% speed-up compared to PDTool, under SSB, TPC-H, TPC-H Skew, TPC-DS and IMDb, respectively. %
It is notable that in Figure~\ref{fig:random-total}, the total workload time of PDTool climbs higher than NoIndex on two occasions, in TPC-DS and IMDb. In IMDb, this is due to the same issue discussed previously under dynamic shifting workloads (due to the optimiser's misestimates, favouring the usage of sub-optimal indices, \eg IMDb Q18).  
While PDTool has a much better execution time than NoIndex under TPC-DS (execution time of 5.3h under PDTool vs 9.3h under NoIndex), due to high recommendation time (5.1 hours, see Table~\ref{tab:full}), PDTool ends up with a higher total workload time. Under these 2 benchmarks (TPC-DS and IMDb), MAB provides over 55\% and 1.5\% performance gain over NoIndex, respectively. 
In Figure~\ref{fig:random-convergence}(a--e), we can see five major spikes for PDTool due to the tuning invocations (in rounds 5, 9, 13, 17, 21).

\begin{table*}[ht]
\begin{center}
\centering
\begin{minipage}{0.70\textwidth}
\caption{Total time breakdown (min): the best choice is in bold text}
\begin{tabular}{|l|l|l|l|l|l|l|l|l|l|} 
\hline
\multicolumn{2}{|l|}{\multirow{2}{*}{Workload}} & \multicolumn{2}{l|}{Recommendation} & \multicolumn{2}{l|}{Creation} & \multicolumn{2}{l|}{Execution} & \multicolumn{2}{l|}{Total} \\ 
\cline{3-10}
\multicolumn{2}{|l|}{} & PDTool & MAB & PDTool & MAB & PDTool & MAB & PDTool & MAB \\ 
\hline
\parbox[t]{2mm}{\multirow{5}{*}{\rotatebox[origin=c]{90}{Static}}} & SSB & 0.34 & \textbf{0.02} & \textbf{0.95} & 1.86 & \textbf{12.9} & 13.15 & \textbf{14.19} & 15.03 \\ 
\cline{2-10}
 & TPC-H & 0.6 & \textbf{0.08} & \textbf{2.45} & 5.66 & \textbf{46.35} & 55.64 & \textbf{49.4} & 61.38 \\ 
\cline{2-10}
 & TPC-H Skew & 0.58 & \textbf{0.11} & \textbf{8.37} & 19.82 & 54.17 & \textbf{32.06} & 63.12 & \textbf{51.99} \\ 
\cline{2-10}
 & TPC-DS & 44.86 & \textbf{1.53} & \textbf{1.45} & 5.94 & 302.63 & \textbf{242.15} & 348.94 & \textbf{249.62} \\ 
\cline{2-10}
 & IMDB & 0.34 & \textbf{0.31} & \textbf{1.1} & 1.3 & 11.01 & \textbf{9.42} & 12.41 & \textbf{11.03} \\ 
\hline
\parbox[t]{2mm}{\multirow{5}{*}{\rotatebox[origin=c]{90}{Dynamic}}} & SSB & 1.28 & \textbf{0.05} & \textbf{1.5} & 2.21 & \textbf{5.42} & 5.69 & 8.2 & \textbf{7.95} \\ 
\cline{2-10}
 & TPC-H & 1.55 & \textbf{0.12} & \textbf{9.36} & 9.74 & 26.35 & \textbf{25.14} & 37.25 & \textbf{35} \\ 
\cline{2-10}
 & TPC-H Skew & 1.65 & \textbf{0.16} & \textbf{14.98} & 20.96 & 85.49 & \textbf{21.44} & 102.11 & \textbf{42.56} \\ 
\cline{2-10}
 & TPC-DS & 11.13 & \textbf{1.66} & \textbf{6.08} & 16.48 & 187.08 & \textbf{155.65} & 204.29 & \textbf{173.79} \\ 
\cline{2-10}
 & IMDB & 3.09 & \textbf{0.29} & \textbf{1.59} & 2.24 & 11.21 & \textbf{7.93} & 15.89 & \textbf{10.46} \\ 
\hline
\parbox[t]{2mm}{\multirow{5}{*}{\rotatebox[origin=c]{90}{Random}}} & SSB & 2.83 & \textbf{0.02} & \textbf{1.77} & 2.37 & 26.59 & \textbf{16.83} & 30.85 & \textbf{19.22} \\ 
\cline{2-10}
 & TPC-H & 7.55 & \textbf{0.08} & 14.68 & \textbf{7.06} & 84.14 & \textbf{80.43} & 106.37 & \textbf{87.57} \\ 
\cline{2-10}
 & TPC-H Skew & 3.3 & \textbf{0.08} & \textbf{31.74} & 34.68 & 48.71 & \textbf{39.44} & 83.75 & \textbf{74.2} \\ 
\cline{2-10}
 & TPC-DS & 310.22 & \textbf{1.4} & \textbf{8.23} & 19.81 & 323.57 & \textbf{227.02} & 642.01 & \textbf{248.24} \\ 
\cline{2-10}
 & IMDB & 14.74 & \textbf{0.28} & 2.72 & \textbf{1.14} & 48.55 & \textbf{14.47} & 66.01 & \textbf{15.89} \\
\hline
\end{tabular}
\label{tab:full}
\end{minipage}\hfill\hspace{0.1em}
\begin{minipage}{0.29\textwidth}
\caption{\emph{Static} workloads under different database sizes (min)}
\begin{tabular}{|l|l|l|l|}
\hline
Workload & SF & PDTool & MAB \\ \hline
\multirow{3}{*}{TPC-H} & 1 & \textbf{2.02} & 2.03 \\ \cline{2-4} 
 & 10 & \textbf{49.4} & 61.38 \\ \cline{2-4} 
 & 100 & 891.01 & \textbf{793.40} \\ \hline
\multirow{3}{*}{\begin{tabular}[c]{@{}l@{}}TPC-H \\ Skew\end{tabular}} & 1 & 4.17 & \textbf{3.83} \\ \cline{2-4} 
 & 10 & 63.12 & \textbf{51.99} \\ \cline{2-4} 
 & 100 & 2640.64 & \textbf{1219.33} \\ \hline
\end{tabular}
\label{tab:db-size}
\end{minipage}
\end{center}
\end{table*}

\subsubsection{The impact of database size} 
\label{sec:db-size-impact}
To examine the impact of database size, we run TPC-H uniform and TPC-H Skew static experiments on SF 1, 10 and 100 databases. As previously discussed, under SF 10, MAB performed better in the case of TPC-H Skew and PDTool performed better on TPC-H (see Table~\ref{tab:db-size}). The impact of sub-optimal index choices is even more evident for larger databases, leading to a huge gap between total workload times of MAB and PDTool for TPC-H Skew (44 hours in the former vs 20 hours in the latter case). In TPC-H, PDTool results in a higher total workload time (14.8 hours vs. 13.2 hours for MAB). 
This is mainly due to sub-optimal optimiser decisions, where the optimiser favours the usage of indices (coupled with nested loops joins) when alternative plans would be a better option. For instance, under the recommended indices from PDTool, some instances of Q5 run longer than 8 minutes (using index nested loops join), where others finish in 1.5 minutes (using a plan based on hash joins). We notice that, with larger database sizes, execution time dominates contributing more than 91\% to the total workload time. We observe faster and more accurate convergence of MAB under larger databases, due to a clear difference between rewards for different arms, highlighting MAB's excellent potential benefits for larger databases. 

\subsubsection{Hypothetical index creation vs actual index creation} \label{sec:exploration-exploitation}

Managing the exploration-exploitation balance under a large number of candidate indices, and an enormous number of combinatorial choices is not trivial. PDTool explores using the ``what-if" analysis, which comes under the tool's recommendation time, whereas MAB explores using index creations. 

\textbf{Cost of hypothetical index creation:} When analysing PDTool's recommendation times, it's evident that average time of a single PDTool invocation grows noticeably with training workload size. As an example, PDTool tuning of the TPC-DS benchmark grows from 2.78 minutes in the dynamic shifting setting (25-query workload) to 62.02 minutes in the dynamic random setting (400-query workload). Furthermore, multiple invocations required in dynamic random and shifting settings aggravate the problem further for PDTool (see Table~\ref{tab:full}). On the other hand, PDTool recommendation time rapidly increases with the complexity of the workloads (\eg under 100-query TPC-H workload, PDTool takes below 8 minutes for recommendation, whereas the same size TPC-DS workload tuning takes more than 45 minutes).

However, MAB recommendation times stay significantly lower and stable despite the workload shifts and changes in complexity or size (see Table~\ref{tab:full}). In all experiments, MAB takes less than 1\% of the total workload time for recommendation, except for IMDb where it takes around 2\% (due to low total workload time and higher number of query templates). More than 80\% of this recommendation time is spent on the initial setup (1\textsuperscript{st} round) and the continuous overhead is negligible.

\textbf{Cost of actual index creation:} While actual execution statistics based search allows the MAB to converge to better configurations, as a down side, MAB spends more time on index creation (see Table~\ref{tab:full}). For instance, under TPC-H and TPC-H Skew static experiments, MAB spends 5.6 and 19.8 minutes on index creation where PDTool only spends 2.4 and 8.3 minutes, respectively. Under skewed data, rewards show more variability which delays the convergence for MAB. This leads to higher exploration and greater creation costs. While MAB is still competitive due to efficient exploration, we consider ways to improve its convergence in future work (see Section~\ref{sec:discussion}).

\textbf{Final verdict:} Comparing the total of recommendation and index creation times (henceforth refereed to as \emph{exploration cost}) between MAB and PDTool presents a clear picture about these two exploration methods. From Table~\ref{tab:full} we can observe that, in most cases (9 out of 15) MAB archives a better exploration cost compared to PDTool. However when the workload is small (\eg dynamic shifting) PDTool tends to perform better. TPC-DS, with the highest number of candidate indices (over 3200 indices) among these benchmarks, provides a great test case for exploration efficiency. Under TPC-DS, MAB exploration cost is significantly lower in shifting and random settings, and marginally higher in the static setting. Despite the efficient exploration, MAB does not sacrifice recommendation quality in any way (better execution costs in 12 out of 15 cases, with significantly better execution costs under all cases of TPC-DS). 

This efficient exploration is promoted by the linear reward-context relationship along with C$^2$UCB's weight sharing (Section \ref{sec:background}), resulting in a small number of parameters to learn. An arm's identity becomes irrelevant and context (Section \ref{sec:bandits}) becomes the sole determining factor of each arm's expected score, which allows MAB to predict the UCB of a newly arriving arm with known context \emph{without} trying it even once.

\begin{figure*}[t]
\centering
\begin{minipage}{0.26\textwidth}
\centering\includegraphics[width=\textwidth]{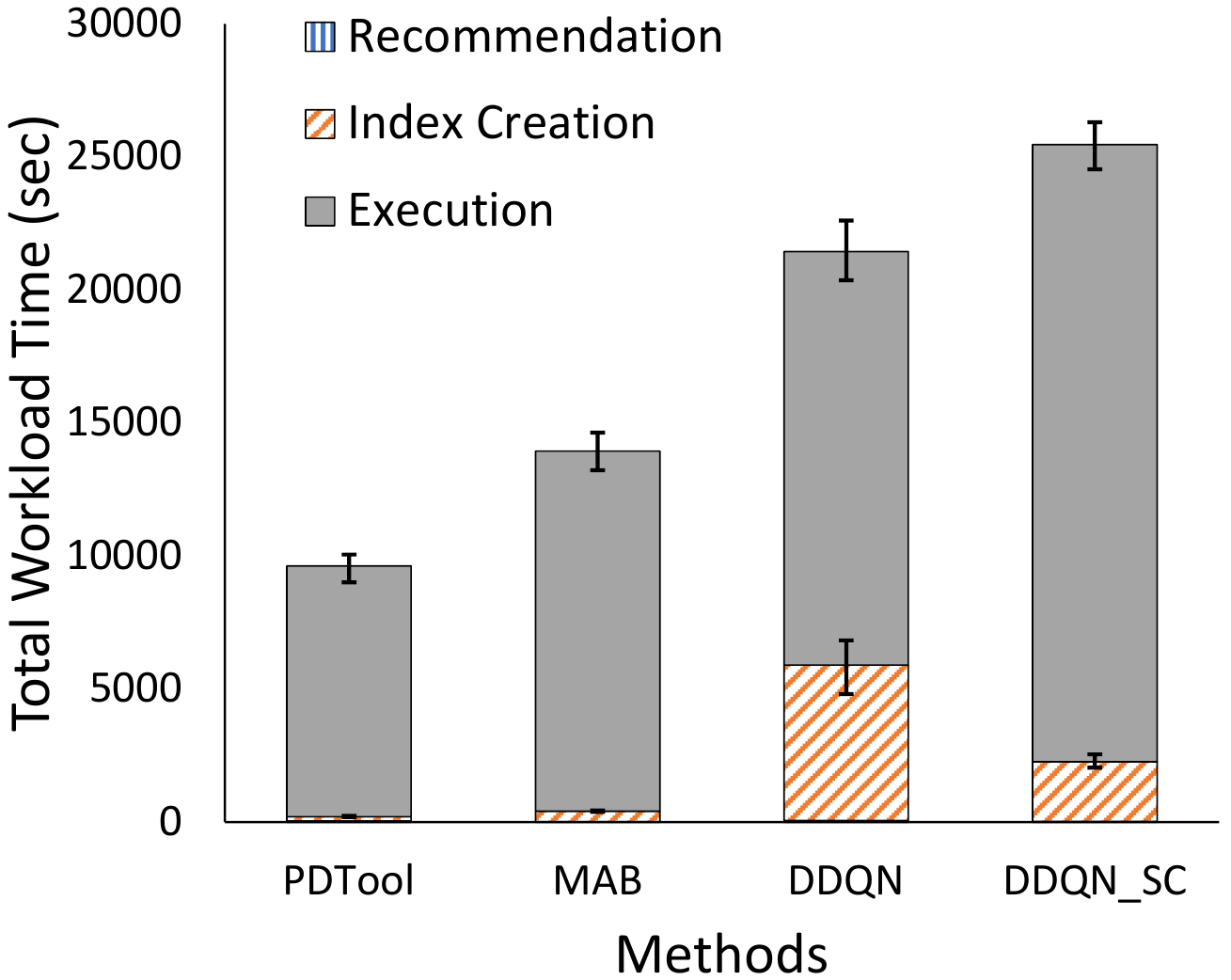}\\ \textbf{(a)}
\end{minipage}\hfill
\begin{minipage}{0.26\textwidth}
\centering\includegraphics[width=\textwidth]{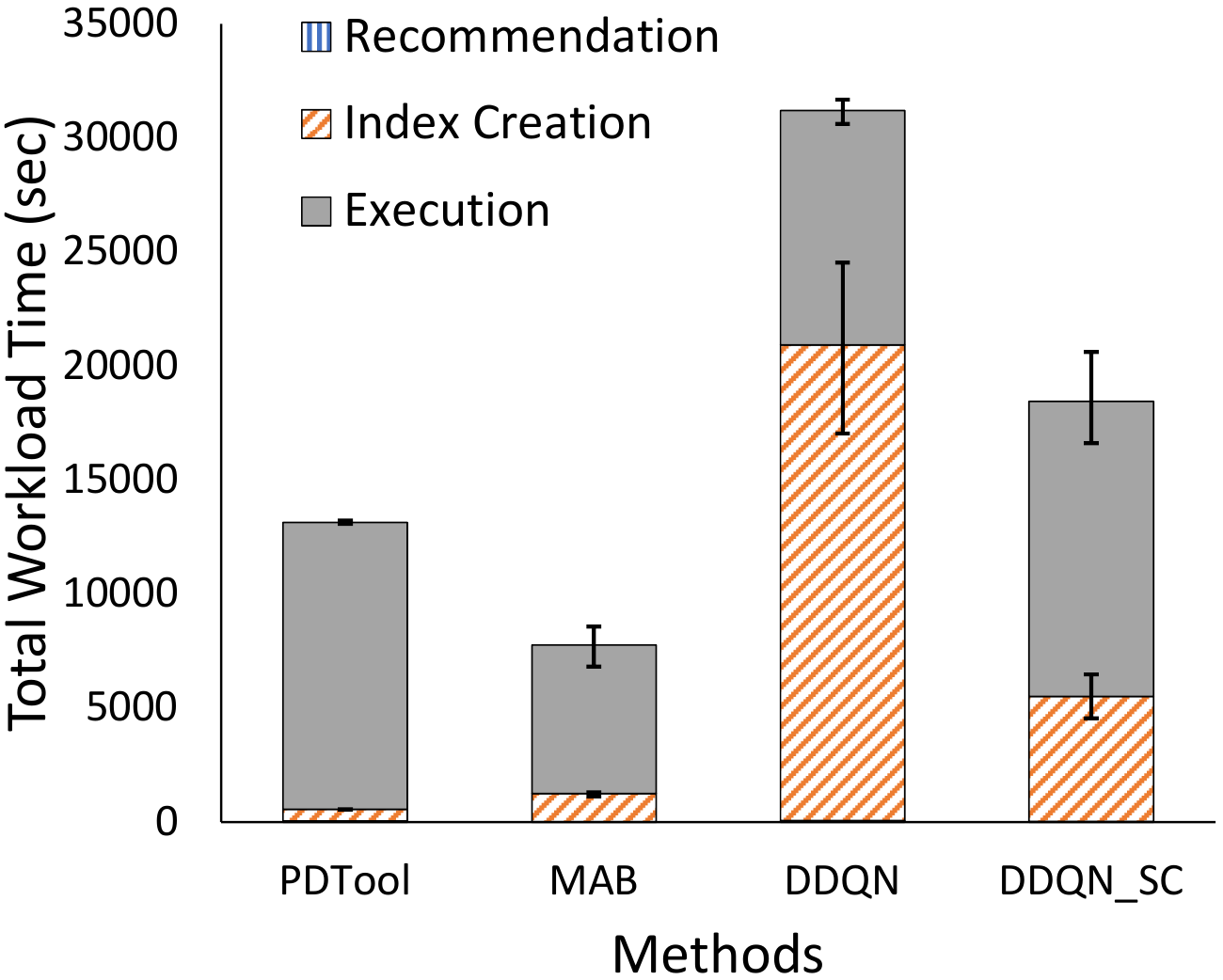}\\ \textbf{(b)}
\end{minipage}\hfill
\begin{minipage}{0.202\textwidth}
\centering\includegraphics[width=\textwidth]{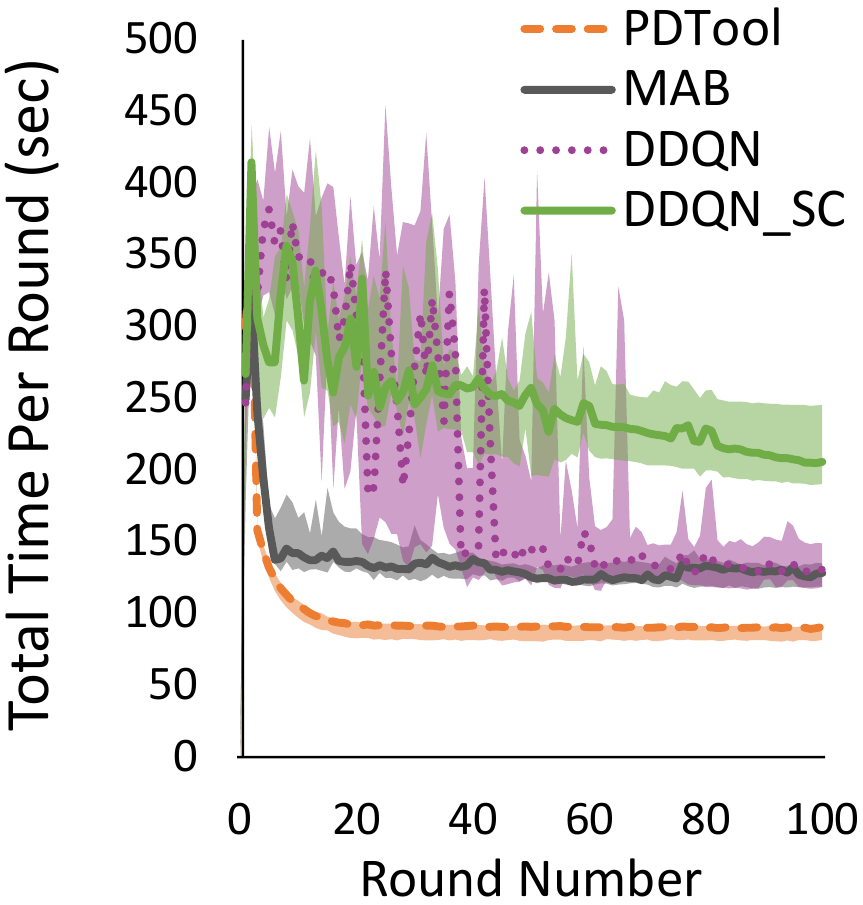}\\ \textbf{(c)}
\end{minipage}\hfill
\begin{minipage}{0.202\textwidth}
\centering\includegraphics[width=\textwidth]{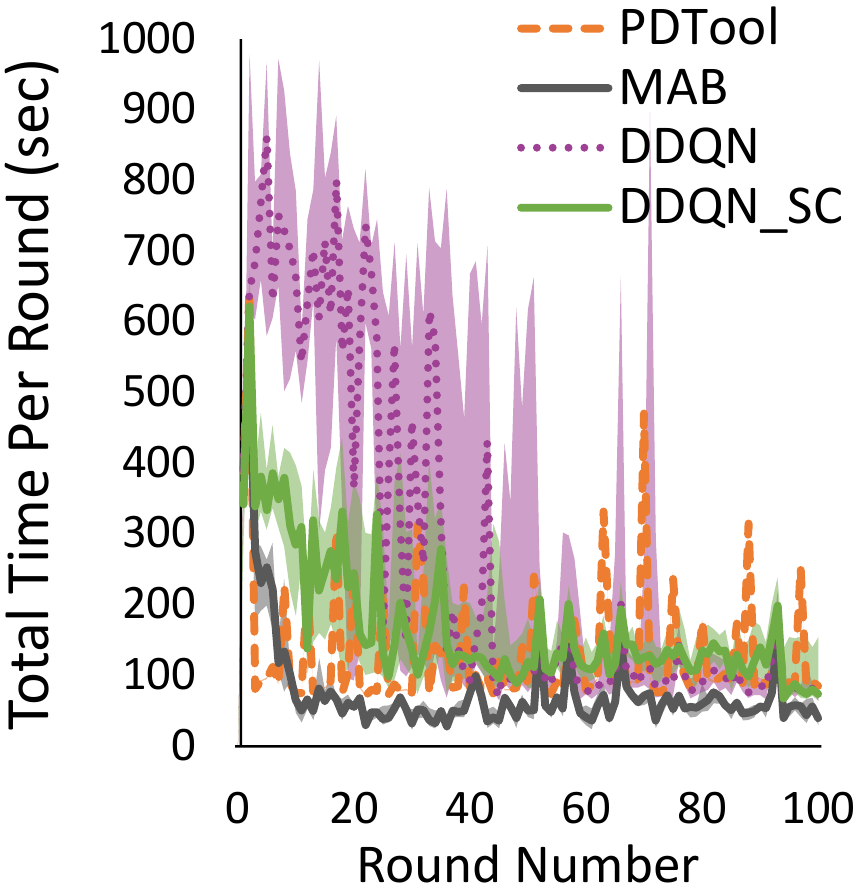}\\ \textbf{(d)}
\end{minipage}\hfill
\\
\caption{DDQN vs. MAB for static workloads: (a) End-to-end workload time for TPC-H, (b) End-to-end workload time for TPC-H Skew, (c) TPC-H convergence, (d) TPC-H Skew convergence.}
\label{fig:rl_comparison}
\end{figure*}

\subsection{Why Not (General) Reinforcement Learning?}\label{sec:not-rl}

Past efforts have considered more general reinforcement learning (RL) for physical design tuning~\cite{no_dba, COREIL}. 
Compared to most MAB approaches, deep RL invites over parameterisation, which can slow convergence (see Figure~\ref{fig:rl_comparison}),
whereas MAB typically provides better convergence, simpler implementation, and \emph{safety guarantees} via strategic exploration and knowledge transfer (see Section~\ref{sec:background}).
Due to its randomisation, RL can also suffer from performance volatility as compared to C$^2$UCB, a deterministic algorithm.

\textbf{Experimental setup.}
The above intuition is supported by experiments with more general RL, where we evaluate the popular DDQN RL agent~\cite{van2016DDQN}.
We run the static 10GB TPC-H and TPC-H Skew benchmark over 100 rounds and present results in Figure~\ref{fig:rl_comparison}.
For a fair comparison, we combine all of MAB's arms' contexts as DDQN state. We also present the same set of candidate indices to the DDQN. For the DDQN's neural network hyperparameters, we followed the experiment of~\cite{no_dba} by setting 4 hidden layers, with 8 neurons each. The discount factor $\gamma$ is set to $0.99$ and the exploration parameter is set to $1$ at the first sample, decaying to 0 with exponential rate reaching 0.01 in the 2400$^\text{th}$ sample. One sample corresponds to one index chosen by the agent. In the beginning of the round, if the agent decides to explore, then the choice of the set of indices will be randomly made for that entire round. %
These experiments are repeated ten times, reporting either average value (Figure \ref{fig:rl_comparison} (a) and (b)) or median ((c) and (d)) along with inter-quartile range. For completeness, we include the case of only using single column indices (DDQN-SC in Figure~\ref{fig:rl_comparison}), as originally proposed  in~\cite{no_dba}.

\textbf{Evaluation.} Due to DDQN-SC's reduced search space in some scenarios, it might not be possible to find the optimal configuration for a workload. This is evident from Figure~\ref{fig:rl_comparison} (a) and (b) where DDQN shows 33\% and 21\% speedup, compared to DDQN-SC, in execution time under TPC-H and TPC-H Skew, respectively. Interestingly, under TPCH-Skew, where the demand for exploration is higher, DDQN-SC has a lower total workload time than DDQN  due to the noticeably small index creation times of single column indices (1.5 hours vs 5.8 hours, respectively). Under both TPC-H and TPC-H Skew, MAB performs significantly better, providing 35\% and 58\% speed-up against the better RL alternative, respectively.

\textbf{No state transitions.}
A strength of more general reinforcement learning is its ability to take into account state transitions---usually represented as transition probability matrices--when actions are taken. However, in the online index selection problem, the importance of state transition is unclear. An approach to  defining state is to think of it as the collection of indices that exist in the system. That is, when an action is taken (\ie choosing an index), it will determine whether that index exists at the start of the next round. This fact, however, does not require a probability estimate since we know with certainty what indices will exist at the beginning of the next round, which by itself determines the next state (\ie deterministic state transition). Another notion of state is the characteristic of the queries asked in the next round. Those queries do not depend on the action taken at the end of the current round, thus it is appropriate to take successor query state as independent of actions taken. Hence, adopting more general RL provides no clear benefit over MAB, while imposing delay to convergence as demonstrated in Figure~\ref{fig:rl_comparison} (c) and (d). The fact that there is no progression of state justifies the choice of using MAB over DNN-based general RL. MABs also enjoy performance guarantees (see Section~\ref{sec:background}).

\textbf{Hyperparameter search space.}
General RL approaches are notorious for challenging  hyperparameter tuning. For example, in this experiment, we have to decide: the number of layers of the neural network, the neurons per layer, activation functions and loss, the exploration parameter $\varepsilon$, and discount factor $\gamma$. Invoking grid search is highly time consuming, considering that one repetition of the experiment for one possible combination of hyperparameters takes hours to complete. In contrast, the C$^2$UCB bandit only requires the hyperparameter $\lambda$---which becomes less relevant as rounds are observed---and $\alpha$ which controls exploration.

\textbf{Volatility of RL.}
Most modern RL algorithms, such as DDQN which we have used here, require randomisation in order to explore vast state-action spaces. There is no hint of which arms or context elements are underexplored, so arms are chosen randomly when the agent decides to enter exploration, which also occurs at random. While we might be lucky and happen upon the optimal set of arms, we could be unfortunate and the algorithm might not encounter useful arms. This is not the case with C$^2$UCB. Extending UCB,  deterministic C$^2$UCB is capable of identifying underexplored arms through their context vectors. The only (rare and as such not strictly necessary) case when  C$^2$UCB is random is where the MAB must tie-break arms. A more significant cause of stability of our MAB is its small parametrisation compared to deep learner-based RL.
Combined, the stable MAB yields a more consistent result, as can be seen in Figure~\ref{fig:rl_comparison}(a) and (b). Much wider variance (highlighted by the error bars) on the DDQN plot demonstrates how the performance of DDQN can vary significantly for the same exact experiments, compared to the narrow error bars on the MAB, which demonstrates the algorithm's stability.

\section{Related Work}
\label{sec:related}
\textbf{Automated physical design tuning.}
Most commercial DBMS vendors nowadays
 offer physical design tools in their products~\cite{DTA2005, DB2_IntegratedApproach, OracleAdvisor}.
These tools rely heavily
on the query optimiser 
to compare benefits of different design
structures without materialisation~\cite{ChaudhuriWhatIf}.
Such an approach is ineffective when base data statistics are unavailable, skewed, or change dynamically~\cite{SurajitDecade}.
In these dynamic environments, the problem of physical design is aggravated: %
a) deciding \emph{when} to call a tuning process is not straightforward;
and b) deciding \emph{what} is a representative training workload is a challenge.

\textbf{Online physical design tuning.} Several research groups have recognised these problems and have offered lightweight solutions to physical design tuning \cite{ToTuneOrNot, OnlineApproachAutoAdmin, QUIET, COLT2}. While such solutions are more flexible and need not know the workload in advance, they are typically limited in terms of applicability to new unknown workloads (generalisation beyond past), and do not come with theoretical guarantees that extend to actual runtime conditions. Moreover, by giving the optimiser a central role, the tools remain susceptible to its mistakes~\cite{AIMeetsAI}. \cite{das2019automatically} extends~\cite{DTA2005} with the use of additional components, in a narrowed scope of index selection to mimic an online tool. This takes corrective actions against the optimiser mistakes through a validation process.

\textbf{Adaptive and learning indices.}
Another dimension of online physical design tuning is database
cracking and adaptive indexing that smooth the creation cost of indices %
by piggybacking on query execution~\cite{StratosCracking, SelfTuningIndexes}. %
Recent efforts have gone a step further and  
proposed replacing data structures with learned models that are smaller in size and faster to query~\cite{learningIndicesKraska, fitingTreeKraska, dataCalculatorStratos}. 
Such approaches are complementary to our efforts: once the data structures (or models) are materialised inside a DBMS, the MAB framework can be used to automate the decision making as to which data structure should be used to speed-up query analysis.

\textbf{Learning approaches to optimisation and tuning.}
Recent years have witnessed new machine learning approaches %
to automate decision-making processes within databases.
For instance, reinforcement learning approaches have been %
used for query optimisation and join ordering~\cite{cuttlefish, trummer2018skinnerdb, deepLearningVictor, handsFreeOlga}. In~\cite{AIMeetsAI}, simpler approaches like regression have mitigated the optimiser's cost misestimates as a path toward more robust index selection. While \cite{AIMeetsAI} shows promising results when avoiding query regressions, according to the authors, when it starts without any knowledge on tuning database (AdaptiveDB), it only provides a marginal workload execution cost improvement (and sometimes deterioration) over the traditional optimiser. Furthermore, this classifier incurs up to 10\% recommendation time, impacting recommendation cost in all cases, especially where recommendation cost already dominates the cost for PDTool (TPC-DS, IMDb). 
When it comes to tuning, the closest approaches employ variants of RL for index selection or partitioning~\cite{no_dba, partitioningCarsten, COREIL} or configuration tuning~\cite{pavlo}.  \cite{COREIL} describes RL-based index selection, which depends solely on the recommendation tool for query-level recommendations and is affected by decision combinatorial explosion, both issues addressed in our work. Unlike its more general counterpart (RL), MABs have advantages of faster convergences as demonstrated in Section~\ref{sec:not-rl}, simple implementation, and theoretical guarantees.  %

\section{Discussion and Future Avenues}
\label{sec:discussion}

This paper scratches the surface of the numerous opportunities for applying bandit learners to performance tuning of databases. In the following, we discuss a rich research vision for the area.

\textbf{Real-world use and integration}.
Even though we use synthetic benchmarks, they are comprehensive in summarising fundamental properties and known pain points of real-life workloads (complexity, data skewness, numerous joins, etc.). Therefore our findings and results generalise to real-life use cases as well. Furthermore, MAB only requires execution statistics to function and can be easily integrated with any existing DBMS.

\textbf{Multi-tenant and HTAP environments}.
A crucial advantage of the MAB setting is theoretical guarantees on the fitness of proposed indices to observed run-time conditions. This is critical for production systems in the cloud and multi-tenant environments~\cite{AIMeetsAI, sqlvm, das2019automatically}, where analytical modelling is impossible due to unpredictable changes in run-time conditions. 
Similar requirements hold for hybrid OLTP/OLAP processing environments (HTAP) where the presence of transactions hinders the usefulness of indices, making analytical modelling next to impossible. 
The MAB approach on the contrary eschews the optimiser and modelling completely, choosing indices based on observed query performance and is thus equally applicable to these challenging environments.

\textbf{Beyond index choices}.
Despite focusing solely on the task of index selection in this paper, the MAB framework is equally applicable to other physical design choices, such as materialised views selection, statistics collection, or even selection of design structures that are a mix of traditional and approximate data structures, such as learned models~\cite{learningIndicesKraska} or other fine-grained design primitives~\cite{dataCalculatorStratos}. 

\textbf{Complementing the recommendation tool}.
Even though the MAB solution was presented as an alternative to the recommendation tool (PDTool), it can work in concert with  existing recommendation systems. MAB can be used as a validator that starts with the PDTool's recommendations and performs further run-time optimisations based on observed performance, in a similar vein to~\cite{das2019automatically}.

\textbf{Cold-start problem.}
Under the current setup, MAB starts without any secondary indices or knowledge about their benefits, forming a cold-start problem and leading to higher creation costs. While MAB is already superior against PDTool even with the creation cost burden (see Section~\ref{sec:exploration-exploitation}), even faster convergence and better creation costs can be provided by pre-training models in hypothetical rounds~\cite{zhang2019warm} (using what-if) or workload forecasting~\cite{forecastingPavlo} to improve context quality.

\textbf{Opportunities for bandit learning.}
The increased search space of possible design choices calls for advancements to bandits algorithms and theory, where unbounded/infinite numbers of arms will be increasingly important. Similarly, various flavours of physical design might ask for novel bandits that adopt the notion of heterogeneous arms (indices, views, or statistics), or hierarchical models where individual choices at lower levels (\eg the choice of indices or materialised views) influence decisions at a higher level (\eg index merging due to memory constraints).

\section{Conclusions}
\label{sec:conclusion}

This paper develops a multi-armed bandit learning framework for online index selection. 
Benefits include eschewing the DBA and the (error-prone) query optimiser by learning the benefits of indices through strategic exploration and observation. 
We justify our choice of MAB over general reinforcement learning for online index tuning, comparing MAB against DDQN, a popular RL algorithm based on deep neural networks, demonstrating significantly faster convergence of the MAB. 
Furthermore, our extensive experimental evaluation demonstrates advantages of MAB over an existing commercial physical design tool (up to 75\% speed up, and 23\% on average), and exemplifies robustness to data skew and unpredictable ad-hoc workloads.

\bibliographystyle{IEEEtran}
\bibliography{IEEEabrv,bibliography}

\end{document}